\documentclass[12pt]{article}
\usepackage{graphicx}
\usepackage{natbib} 
\usepackage{url} 
\usepackage{amsmath} 
\usepackage{amsfonts}
\usepackage{amssymb} 
\usepackage{bbm}
\usepackage{lineno}
\usepackage{setspace}
\usepackage{lscape} 
\usepackage{float} 
\usepackage{breakcites}
\usepackage{pdfpages}
\usepackage{xcolor}
\usepackage[colorlinks=true,citecolor=black,linkcolor=black,hidelinks]{hyperref}
\usepackage{setspace}
\usepackage{authblk}
\usepackage{makecell}
\usepackage{multirow}
\usepackage{booktabs}
\usepackage[title]{appendix}

\graphicspath{{Figures/}}

\newcommand{\blind}{0}
\newcommand{\bs}{\boldsymbol}
\newcommand{\bx}{\bs{x}}
\newcommand{\bu}{\bs{u}}

\newcommand{\by}{\bs{y}}

\newcommand{\bX}{\bs{X}}

\newcommand{\bw}{\bs{w}}
\newcommand{\bth}{\bs{\theta}}

\usepackage{ntheorem}
\makeatletter
\renewtheoremstyle{plain}
  {\item{\theorem@headerfont ##1\ ##2\theorem@separator}~}
  {\item{\theorem@headerfont ##1\ ##2\ (##3)\theorem@separator}~}
\makeatother

\newtheorem{definition}{Definition}
\newtheorem{assumption}{Assumption}
\newtheorem{theorem}{Theorem}
\newtheorem{proposition}{Proposition}

\begin{document}

\def\spacingset#1{\renewcommand{\baselinestretch}%
{#1}\small\normalsize} \spacingset{1}

\if0\blind
{
  \title{\bf Measuring Sample Quality in Algorithms for Intractable Normalizing Function Problems}
    \author[1]{Bokgyeong Kang}
    \author[2]{John Hughes}
    \author[3]{Murali Haran}
    \affil[1]{Department of Statistical Science, Duke University}
    \affil[2]{College of Health, Lehigh University}
    \affil[3]{Department of Statistics, Pennsylvania State University}
    \date{}
  \maketitle
} \fi

\if1\blind
{
  \bigskip
  \bigskip
  \bigskip
  \begin{center}
    {\LARGE\bf Measuring Sample Quality in Algorithms for Intractable Normalizing Function Problems}
    \end{center}
  \medskip
} \fi

\begin{abstract}
Models with intractable normalizing functions have numerous applications. Because the normalizing constants are functions of the parameters of interest, standard Markov chain Monte Carlo cannot be used for Bayesian inference for these models. A number of algorithms have been developed for such models. Some have the posterior distribution as their asymptotic distribution. Other ``asymptotically inexact'' algorithms do not possess this property. There is limited guidance for evaluating approximations based on these algorithms. Hence it is very hard to tune them. We propose two new diagnostics that address these problems for intractable normalizing function models. Our first diagnostic, inspired by the second Bartlett identity, is in principle broadly applicable to Monte Carlo approximations beyond the normalizing function problem. We develop an approximate version of this diagnostic that is applicable to intractable normalizing function problems. Our second diagnostic is a Monte Carlo approximation to a kernel Stein discrepancy-based diagnostic introduced by Gorham and Mackey (2017). We provide theoretical justification for our methods and apply them to several algorithms in challenging simulated and real data examples including an Ising model, an exponential random graph model, and a Conway--Maxwell--Poisson regression model, obtaining interesting insights about the algorithms in these contexts.
\end{abstract}

\noindent%
{\it Keywords:} Bartlett identity, Doubly intractable distributions, Kernel Stein discrepancy, Markov chain Monte Carlo, Sample quality measure
\vfill

\newpage
\spacingset{1.5} 


\section{Introduction}
\label{sec:intro}


Models with intractable normalizing functions arise in many contexts, notably the Ising \citep{lenz1920,ising1925} and autologistic models \citep[see][for a review]{Besag1974,Hughes2011autologistic} for binary data on a lattice, exponential random graph models \citep[see][]{Robins2007,Hunter2006} and mixed graphical models \citep[see][]{Lauritzen1989,Lee2015,Cheng2017} for explaining relationships among actors in networks, interaction point process models \citep[see][]{strauss1975,Goldstein2015} for describing spatial patterns among points, and Conway--Maxwell-Poisson regression models \citep[see][]{Conway1962,Shmueli2005,Chanialidis2018} for over- or under-dispersed count data. 
Interest in models having intractable normalizing functions has increased rapidly during the last two decades. Indeed, a look at the Web of Science database shows that the number of yearly citations of articles on this subject has grown quadratically since 2005. Unfortunately, standard Markov chain Monte Carlo (MCMC) algorithms, the workhorse of Bayesian inference in the past few decades, cannot be applied to these models. Suppose we have data $\bx \in \mathcal{X}$ generated from a probability model $f(\bx \mid \bth)$ with likelihood function $L(\bth \mid \bx) = h(\bx \mid \bth)/c(\bth)$, where $c(\bth)$ is a normalizing function, and a prior density $p(\bth)$. The posterior density of $\bth$ is $\pi(\bth \mid \bx) \propto p(\bth) h(\bx \mid \bth)/c(\bth)$, which brings about so-called doubly intractable posterior distributions. A major problem in constructing a standard MCMC algorithm for such models is that $c(\bth)$ cannot be easily evaluated. The Metropolis--Hastings (MH) algorithm \citep{Metropolis1953,Hastings1970} acceptance probability at each step requires evaluating the unknown ratio $c(\bth)/c(\bth^\prime)$, where $\bth^\prime$ denotes the proposed value.

A wide range of computational methods have been proposed for Bayesian inference for doubly intractable posterior distributions \citep[see][for a review]{Park2018}. There are asymptotically exact algorithms whose Markov chain has a stationary distribution equal to its target distribution \citep[cf.][]{moller2006,murray2006,atchade2013,Lyne2015,Liang2016}. Throughout this manuscript we use ``target distribution'' to refer to the posterior distribution of interest. 
Some of the asymptotically exact algorithms are available only for a small class of probability models having intractable normalizing functions. The other algorithms are complicated to construct and have to be carefully tuned to provide reliable inference. These algorithms tend to be computationally intensive \citep{Park2018}. There are asymptotically inexact algorithms that may be much faster and can be applied to a wider class of problems \citep[cf.][]{Liang2010,alquier2016,Park2020}. An asymptotically inexact algorithm either converges to an approximation of the target or is not known to converge to any distribution. The performance of these algorithms relies on the choice of various tuning parameters. It is also not always easy to judge the tradeoffs between using a faster asymptotically inexact algorithm and a potentially more computationally expensive but asymptotically exact algorithm. Hence, it is crucial to have diagnostics that provide guidance for users to carefully tune their algorithms to provide reliable results. 

There is a large literature on convergence diagnostics for MCMC algorithms \citep[see][for a review]{Cowles1996,Flegal2011,Roy2020}. In fact, given the wide usage of MCMC and the importance of MCMC diagnostics, some of the best known MCMC diagnostics papers have thousands of citations \citep[cf.][]{gelman1992inference,geyer1992practical,geweke1991evaluating} and at this point perhaps often even get used without citation. However, the literature on assessing the quality of approximations provided by asymptotically inexact algorithms is very limited. There are several approaches that measure the deviation between sample means and target expectations whose values are known \citep{Fan2006,Gorham2015,Gorham2017}. \citet{Lee2019} and \citet{Xing2019} have provided tools for assessing the coverage of approximate credible intervals. These are laudable innovations, but they are not available for asymptotically inexact algorithms for doubly intractable distributions. This motivates our development of sample quality measures that assist scientists in tuning these algorithms.

In this article we describe a new diagnostic method that uses the well known second Bartlett identity \citep{Bartlett1953a,Bartlett1953b}. Our method is, in principle, applicable in virtually any likelihood-based context where misspecification is of concern. We develop a Monte Carlo approximation to this new diagnostic that is applicable to intractable normalizing function models. We also develop an approximate version of the kernel Stein discrepancy introduced by \citet{Gorham2017}, making this available for doubly intractable distributions. This diagnostic asymptotically inherits the same convergence properties as that of \citet{Gorham2017} and thus can be used for diagnosing convergence of a sequence of sample distributions to the target distribution. Following \citet{Gorham2017}, we think of our diagnostics as measuring ``sample quality.''

The remainder of this article is organized as follows. In Section~\ref{sec:intractable normalizing functions} we briefly describe computational methods for models with intractable normalizing functions. In Section~\ref{sec:dagnostics} we discuss the need for diagnostics for tuning asymptotically inexact algorithms. In Section~\ref{sec:CD} we propose a new diagnostic for asymptotically exact and inexact methods, and we develop an approximation for doubly intractable distributions. In Section~\ref{sec:KSD} we briefly describe the kernel Stein discrepancy introduced by \citet{Gorham2017} and propose its Monte Carlo approximation for intractable normalizing function models. We provide theoretical justification for our diagnostics. In Section~\ref{sec:app} we describe the application of our diagnostic approaches to several algorithms in the context of three different challenging examples and study the computational complexity and variability of our diagnostics. Finally, in Section~\ref{sec:discussion} we conclude with a brief summary and discussion.

\section{Inference for Models with Intractable Normalizing Functions}
\label{sec:intractable normalizing functions}

Several computational methods have been developed for Bayesian inference for models with intractable normalizing functions.
\citet{Park2018} categorized these algorithms into two general, overlapping classes: (1) \textit{auxiliary variable approaches}, which introduce an auxiliary variable and cancel out the normalizing functions in the Metropolis--Hastings acceptance probability \citep{moller2006,murray2006, Liang2010, Liang2016}, and (2) \textit{likelihood approximation approaches}, which directly approximate the normalizing functions and plug the approximations into the Metropolis--Hastings acceptance probability \citep{atchade2013, Lyne2015, alquier2016, Park2020}. An important characteristic of these algorithms is whether they are ``asymptotically exact'' or ``asymptotically inexact.'' Asymptotically exact algorithms generate a sequence whose asymptotic distribution is exactly equal to the target distribution. Asymptotically inexact algorithms generate a sequence that does not converge to the target distribution (or to any distribution in some cases).

Asymptotically exact algorithms have attractive theoretical properties but can often be computationally burdensome or even infeasible for challenging models. For instance, \citet{moller2006} and \citet{murray2006} depend on perfect sampling \citep[cf.][]{propp1996}, an algorithm that generates an auxiliary variable exactly from the target distribution using bounding Markov chains. Perfect samplers tend to be very computationally expensive and are available only for a small class of probability models. \citet{atchade2013} and \citet{Liang2016} propose asymptotically exact algorithms that do not need perfect sampling. \citet{atchade2013}'s adaptive MCMC (ALR) algorithm introduces multiple particles $\bth^{(1)}, \dots, \bth^{(d)}$ over the parameter space and approximates $c(\bth)$ in the acceptance probability through importance sampling using the entire sample path of the algorithm. \citet{Liang2016}'s adaptive exchange (AEX) algorithm runs an auxiliary chain and a target chain simultaneously. At each iteration, the auxiliary chain simulates and stores a sample from a set of distributions, $\{h(\bx \mid \bth^{(1)})/c(\bth^{(1)})$, $\dots$, $h(\bx \mid \bth^{(d)})/c(\bth^{(d)})\}$, where $\bth^{(1)}$, $\dots$, $\bth^{(d)}$ are predetermined particles over the parameter space. The target chain generates a posterior sample via the exchange algorithm \citep{murray2006}, where an auxiliary variable $\by$ is sampled from the cumulative samples in the auxiliary chain instead of exact sampling of $\by$. These algorithms require storing simulated auxiliary data or their sufficient statistics with each iteration. The computational and memory costs are very expensive for models without low-dimensional sufficient statistics. Pseudo-marginal MCMC algorithms \citep{Beaumont2003,Andrieu2009} are approaches that use an unbiased Monte Carlo approximation of an intractable likelihood. \citet{Lyne2015} constructed an unbiased likelihood estimate for models with intractable normalizing functions. To obtain a single estimate, their approach requires multiple Monte Carlo approximations to the normalizing constant and thus can often be computationally prohibitive. 

Several computationally efficient asymptotically inexact algorithms have also been proposed. For instance, the double Metropolis--Hastings (DMH) sampler, proposed by \citet{Liang2010}, replaces perfect sampling with a standard Metropolis--Hastings algorithm, an ``inner sampler,'' at each iteration of the exchange algorithm. The DMH algorithm is easy to implement and is computationally efficient compared to the other algorithms discussed thus far. But the inner sampling becomes more computationally expensive with an increase in the dimension of the data. For large data problems, \citet{Park2020} proposed a function emulation algorithm that approximates the likelihood normalizing function (or full likelihood function) at several parameter values and interpolates the function using a Gaussian process. This provides significant gains in computational efficiency. More asymptotically exact and inexact algorithms are found in \citet{Park2018}.

Both asymptotically exact and inexact algorithms require careful tuning in order to provide good approximations in a reasonable amount of time. 
For instance, the ALR and AEX algorithms require users to select an appropriate number $d$ of particles to cover the important region of the parameter space. The DMH algorithm requires users to decide the length $m$ of the inner sampler for generating an auxiliary variable. As $m$ grows large the auxiliary variable becomes approximately a draw from the probability model at the expense of longer computing time. However, currently there is little guidance on how to tune these algorithms, and most of them rely on simulation studies. Also, the behavior of the algorithm varies across models or across datasets for a given model.

\section{The Need for Diagnostics for Intractable Normalizing Function Problems}
\label{sec:dagnostics}

There is a vast literature on MCMC convergence diagnostics \citep[see][for a review]{Cowles1996,Flegal2011,Roy2020}. However, these diagnostics may not be adequate for asymptotically inexact algorithms. Suppose we have a sample generated by an asymptotically inexact method. Such a sample may not have an asymptotic distribution, or said sample may converge but to a mere approximation of the target distribution. Standard MCMC diagnostics assess whether the sample has converged to its asymptotic distribution but do not assess whether the asymptotic distribution is equal to the target distribution. As discussed in Section~\ref{sec:intractable normalizing functions}, however, asymptotically exact algorithms for models with intractable normalizing functions are available only for a few special cases, and even for these cases computing tends to be quite burdensome. 

Several approaches based on measuring the difference between sample means and target expectations have been proposed for assessing the quality of approximations provided by asymptotically inexact algorithms. \citet{Fan2006} proposed score function (i.e., gradient of the log likelihood) diagnostics for assessing estimates of some quantities, the values of which are known under the target distribution. They suggested plotting the Monte Carlo estimate versus the sample size together with error bounds. \citet{Gorham2015} pointed out limitations of the score function diagnostics caused by considering only a finite class of functions and introduced a new diagnostic method based on a Stein discrepancy. \citet{Gorham2015} defined Stein discrepancies that bound the discrepancy between the sample mean and the target expectation over a large collection of functions whose target expectations are zero. The Stein discrepancies are supported by a theory of weak convergence and are attainable by solving a linear program. Combining this idea with the theory of reproducing kernels, \citet{Gorham2017} provided a closed-form kernel Stein discrepancy with sound theoretical support analogous to that of \citet{Gorham2015} (see Section~\ref{subsec:IMQKSD} for details). These approaches are useful for comparing asymptotically approximate samplers and for selecting tuning parameters for such samplers. However, all of them require evaluating the score function of the target density, which is not possible for doubly intractable posterior distributions. In contrast, our approaches apply broadly to asymptotically exact and inexact algorithms even for such challenging problems. To our knowledge, no other diagnostics are currently available for asymptotically inexact algorithms for intractable normalizing function models. We have studied our diagnostics as applied in several challenging real data settings. In addition, we provide theoretical justification for our methods.

\section{Curvature Diagnostics}
\label{sec:CD}
In this section we introduce two new diagnostics that are useful for tuning asymptotically exact and inexact algorithms: a curvature diagnostic (CD) and an approximate curvature diagnostic (ACD). The curvature diagnostic is based on the second Bartlett identity from the classical theory of maximum likelihood. The approximate curvature diagnostic is an approximation of the CD that is suitable for intractable normalizing function problems.

\subsection{A General Purpose Curvature Diagnostic}
\label{subsec:CD}

Consider a sequence of sample points $\bth^{(1)}, \dots, \bth^{(n)}$ generated by an algorithm having $\pi(\bth \mid \bx)$ as its target distribution. Our aim is to determine whether the sample will produce a good approximation to some quantity of interest, e.g., $\textup{E}_{\pi}\{z(\bth)\}$, where $z(\bth)$ is a real-valued function. Our diagnostic, the curvature diagnostic, is inspired by the method for obtaining the asymptotic variance of a maximum likelihood estimator under a misspecified model. When the model is misspecified, the second Bartlett identity does not hold, which is to say (see details below) the sensitivity matrix is not equal to the variability matrix. And so the asymptotic variance of the estimator does not simplify to the inverse of the Fisher information. Our curvature diagnostic uses the dissimilarity between the sensitivity matrix and the variability matrix to assess the quality of the sample. We provide details in the following paragraph.

Let $\bs{u}(\bth) = \nabla_{\bth} \log \pi(\bth \mid \bx)$ be the score function of the posterior density $\pi(\bth \mid \bx)$. The posterior density has the sensitivity matrix
\begin{align*}
    \textup{E}_{\pi} \left\{ - \frac{\partial}{\partial \bth} \bs{u}(\bth) \right\} &= \int_{\bs{\Theta}}  - \frac{\partial}{\partial \bth} \bs{u}(\bth) \pi(\bth \mid \bx) d\bth
\end{align*}
and the variability matrix $\textup{var}_{\pi} \left\{ \bs{u}(\bth) \right\} = \textup{E}_{\pi} \left\{ \bs{u}(\bth) \bs{u}(\bth)^\top \right\}$.
Note that the identity for the variability matrix follows from Bartlett's first identity: $\textup{E}_{\pi}\{\bs{u}(\bth)\} = \textbf{0}$. Let $H(\bth) = \frac{\partial}{\partial \bth} \bs{u}(\bth)$, $J(\bth) = \bs{u}(\bth) \bs{u}(\bth)^\top$, and $\bs{d}(\bth) = \textup{vech}[J(\bth) + H(\bth)]$, where $\textup{vech}(M)$ denotes the half-vectorization of the matrix $M$. 
We have $\textup{E}_{\pi}\{\bs{d}(\bth)\} = \textbf{0}$ by Bartlett's second identity: $\textup{E}_{\pi}\{-H(\bth)\} = \textup{E}_{\pi}\{J(\bth)\}$. Using the sample $\bth^{(1)}, \dots, \bth^{(n)}$ we construct a Monte Carlo approximation to the half-vectorized difference between the sensitivity and variability matrices as
\begin{align}
 \bs{d}_n &:= \frac{1}{n} \sum_{i=1}^n \bs{d}(\bth^{(i)}). \label{eq:dbarn}
\end{align}

Suppose the samples $\bth^{(1)}, \dots, \bth^{(n)}$ are independent. By the central limit theorem, we have
\begin{align}
    \sqrt{n} \bs{d}_n &\stackrel{d}{\to} \textup{N}(\textbf{0}, V), \nonumber
\end{align}
where $V := \textup{cov}_{\pi}\{\bs{d}(\bth)\} = \textup{E}_{\pi} \left\{ \bs{d}(\bth) \bs{d}(\bth)^\top \right\}$. Our unbiased and consistent approximation of $V$ is calculated as
\begin{align}
    V_n &:= \frac{1}{n} \sum_{i=1}^n \bs{d}(\bth^{(i)})\bs{d}(\bth^{(i)})^\top. \label{eq:Vhat}
\end{align}
If $\bth^{(1)},\dots, \bth^{(n)}$ are from a Markov process and thus dependent, then by the Markov chain Monte Carlo central limit theorem we have
\begin{align*}
    \sqrt{n} \bs{d}_n \stackrel{d}{\to} \textup{N}(\textbf{0}, \Sigma),
\end{align*}
where $\Sigma := \textup{cov}_{\pi} \{ \bs{d}(\bth^{(i)})\} + 2 \sum_{k=1}^{\infty} \textup{cov}_{\pi}\{ \bth^{(i)}, \bth^{(i+k)}\}$. The asymptotic covariance matrix $\Sigma$ can be estimated by the multivariate batch means method \citep{Chen1987,Flegal2011}. Let $n = a_n b_n$ where $a_n$ is the number of batches and $b_n$ is the batch size. The batch means estimate of $\Sigma$ is calculated as
\begin{align}
    \Sigma_n := \frac{b_n}{a_n-1} \sum_{j = 1}^{a_n} (\Bar{\bs{d}}^j_{b_n} - \bs{d}_n)(\Bar{\bs{d}}^j_{b_n} - \bs{d}_n)^\top, \label{eq:Sigmahat}
\end{align}
where $\Bar{\bs{d}}^j_{b_n} = \sum_{i = (j-1)b_n + 1}^{jb_n} \bs{d}(\bth^{(i)})$. The batch means estimator is strongly consistent under some conditions \citep{Damerdji1994,Jones2006,Vats2019}. Because it is simple to construct and appears to work well in practice under quite a wide range of settings \citep[cf.][]{flegal2008markov}, we suggest using batch means. Our curvature diagnostic is then defined as follows.

\begin{definition}[Curvature Diagnostic (CD)]
Consider a sample $\bth^{(1)}$, $\dots$, $\bth^{(n)}$ generated by an algorithm having $\pi(\bth\mid\bx)$ as its target. If the sample is independent, our curvature diagnostic is defined as $\mathcal{C}_n(\bx) := n \bs{d}_n^\top V_n^{-1} \bs{d}_n$. If the sample is from a Markov process, our curvature diagnostic is defined as $\mathcal{C}^{\textup{BM}}_n(\bx) := n \bs{d}_n^\top \Sigma_n^{-1} \bs{d}_n$.
\label{def:CD}
\end{definition}
The CD has an asymptotic $\chi^2(r)$ distribution, where $r = p(p+1)/2$ and $p$ is the dimension of $\bth$, if the asymptotic distribution of the sample is equal to the target distribution.
A simple and effective heuristic for determining $n$ is to plot $\hat{V}_{n}$ against the posterior sample size $n$ and select $n$ at which the approximations appear to have stabilized. Given $n$, an abnormally large value of CD with respect to $\chi^2(r)$ can signal poor sample quality. We use the $1-\alpha$ quantile of the $\chi^2(r)$ as a threshold for our diagnostic. A sample path for which the CD value is below the threshold is considered to have an asymptotic distribution that is reasonably close to the target distribution. We found that $\alpha = 0.01$ performed well in our simulation experiments, and so we take $\alpha$ = 0.01 in the sequel.

We note that our curvature diagnostic is, in principle, applicable not only in intractable normalizing function problems but in virtually any likelihood-based context where misspecification is of concern. Moreover, the framework we present in this article can be extended to incorporate higher moments.

\subsection{An Approximate Curvature Diagnostic for Intractable Normalizing Function Problems}
\label{subsec:ACD}

If the normalizing function $c(\bth)$ of the likelihood is intractable, it is not possible to evaluate the curvature diagnostic since the diagnostic involves $\nabla_{\bth} \log c(\bth)$ and $ \partial \nabla_{\bth} \log c(\bth) / \partial \bth$. 
The intractable terms can be written as
\begin{align}
    \nabla_{\bth} \log c(\bth) &= \textup{E}_{\bx \mid \bth} \left\{  \nabla_{\bth} \log h(\bX \mid \bth) \right\} \label{eq:acdExp1}\\
    \frac{\partial \nabla_{\bth} \log c(\bth)}{\partial \bth} &=\textup{E}_{\bx \mid \bth} \left\{ 
    \frac{\partial \nabla_{\bth} \log h(\bX | \bth)}{\partial \bth} \right\}+ \textup{E}_{\bx \mid \bth} \left\{ [\nabla_{\bth} \log h(\bX | \bth)] [\nabla_{\bth} \log h(\bX | \bth)]^\top \right\} \nonumber \\ 
    & \quad -  \textup{E}_{\bx \mid \bth} \left\{ \nabla_{\bth} \log h(\bX | \bth) \right\}  \textup{E}_{\bx \mid \bth} \left\{ \nabla_{\bth} \log h(\bX | \bth) \right\}^\top. \label{eq:acdExp2}
\end{align}
We can replace the expectations with their Monte Carlo approximations using auxiliary variables $\by^{(1)}, \dots, \by^{(N)}$ that are generated exactly from $f(\cdot \mid \bth)$ or generated by a Monte Carlo algorithm having $f(\cdot \mid \bth)$ as its target distribution. We provide details in Appendix \ref{app:sec:mcapprox}. 

To reduce computational cost, we employ self-normalizing importance sampling \citep[SNIS;][]{Tan2020}. For $i = 1,\dots,n$ we need to estimate $\text{E}_{\bx \mid \bth^{(i)}}\{g(\bX)\}$ where $g$ denotes any function inside of the expectations in \eqref{eq:acdExp1} and \eqref{eq:acdExp2}. For some $\bs{\psi} \in \bs{\Theta}$, the expectation can be written as 
\begin{align*}
    \text{E}_{\bx\mid \bth^{(i)}}\{g(\bX)\}&=\text{E}_{\bx\mid \bs{\psi}} \left\{ g(\bX) \frac{h(\bX \mid \bth^{(i)})}{h(\bX \mid \bs{\psi})} \right\} \Big/ \text{E}_{\bx\mid \bs{\psi}} \left\{ \frac{h(\bX \mid \bth^{(i)})}{h(\bX \mid \bs{\psi})} \right\}.
    \end{align*}
The SNIS introduces a finite set $S = \{\bs{\psi}_1, \dots, \bs{\psi}_M \}$ of particles over the parameter space $\bs{\Theta}$ and simulates auxiliary variables $\by_r^{(1)}, \dots, \by_r^{(N)}$ from $f(\cdot \mid \bs{\psi}_r)$ for each $r = 1, \dots, M$. For $i = 1,\dots,n$, the consistent estimate of $\text{E}_{\bx \mid \bth^{(i)}}\{g(\bX)\}$ can be obtained by $\sum_{j=1}^N w_r^{(j)} g(\by^{(j)}_r)$, where
\begin{align*}
    w_r^{(j)} &= \frac{h(\by^{(j)}_r \mid \bth^{(i)})}{h(\by^{(j)}_r \mid \bs{\psi}_r)} \Big/ \sum_{j^\prime = 1}^N  \frac{h(\by^{(j^\prime)}_{r} \mid \bth^{(i)})}{h(\by^{(j^\prime)}_{r} \mid \bs{\psi}_r)},
\end{align*}
and $\bs{\psi}_r$ is the particle nearest to $\bth^{(i)}$. For particle sampling we use a quasi-random number generator \citep{Faure2009}. The quasi-random number generation can be done easily using R package \texttt{qrng}. We use Mahalanobis distance to measure closeness, and we estimate the covariance matrix using the samples $\bth^{(1)}, \dots, \bth^{(n)}$. We found that $M = 200p$ and $N = 10,000$ performed well in our simulation experiments.

By plugging these approximations into $\bs{d}(\bth)$ we obtain its approximation $\hat{\bs{d}}_N(\bth)$. By replacing $\bs{d}(\bth)$ with its approximation $\hat{\bs{d}}_N(\bth)$ in \eqref{eq:dbarn}, \eqref{eq:Vhat}, and \eqref{eq:Sigmahat} we obtain $\hat{\bs{d}}_{n,N}$, $\hat{V}_{n,N}$, and $\hat{\Sigma}_{n,N}$. We note that the number $a_{n,N}$ of batches and the length $b_{n,N}$ of each batch depend on $n$ and $N$ for $\hat{\Sigma}_{n,N}$.
Then an approximate version of the curvature diagnostic for intractable normalizing function problems can be defined as follows.

\begin{definition}[Approximate Curvature Diagnostic (ACD)]
Consider a sample $\bth^{(1)}, \dots, \bth^{(n)}$ generated by an algorithm having $\pi(\bth\mid\bx)$ as its target. If the sample is independent, our approximate curvature diagnostic is defined as $\mathcal{\hat{C}}_{n, N}(\bx) := n \: \hat{\bs{d}}_{n,N}^\top \hat{V}_{n,N}^{-1} \hat{\bs{d}}_{n,N}$.
If the sample is from a Markov process, our approximate curvature diagnostic is defined as $\mathcal{\hat{C}}^{\textup{BM}}_{n, N}(\bx) := n \: \hat{\bs{d}}_{n,N}^\top \hat{\Sigma}_{n,N}^{-1} \hat{\bs{d}}_{n,N}$.
\label{def:ACD}
\end{definition}

To provide theoretical justification for $\mathcal{\hat{C}}_{n, N}(\bx)$, we make the following assumptions regarding the prior density $p(\bth)$ and the unnormalized probability model density $h(\bx\mid\bth)$. We use $\Vert \cdot \Vert_{\max}$ to represent the max norm of vectors or matrices.
\begin{assumption}
$\Vert \nabla_{\bth} \log p(\bth) \Vert_{\max} < \infty$ and $\Vert \frac{\partial}{\partial \bth} \nabla_{\bth} \log p(\bth) \Vert_{\max} < \infty$.
\label{ass:prior}
\end{assumption}
\begin{assumption}
$\Vert \nabla_{\bth} \log h(\bx \mid \bth) \Vert_{\max} < \infty$ and $\Vert \frac{\partial}{\partial \bth} \nabla_{\bth} \log h(\bx \mid \bth) \Vert_{\max} < \infty$.
\label{ass:likelihood}
\end{assumption}
Since the prior density is determined by the user, the prior's score function and Hessian matrix may be assumed to be bounded. In many applications Assumption~\ref{ass:likelihood} may be checked easily. In particular, Assumption~\ref{ass:likelihood} is satisfied with high probability for exponential families. For a probability model in an exponential family, the model's score is its summary statistics and the first inequality of Assumption~\ref{ass:likelihood} is satisfied with high probability \citep{Chazottes2007}. The model's Hessian matrix is zero and the second inequality is satisfied almost surely.
Under these assumptions, Theorem~\ref{thm:ACD} quantifies the distance between the asymptotic covariance matrix $V$ and its two-stage approximation $\hat{V}_{n,N}$.
\begin{theorem}
Consider an i.i.d. sample $\bth^{(1)}, \dots, \bth^{(n)}$ the asymptotic distribution of which is $\pi(\bth \mid \bx)$. If Assumptions~\ref{ass:prior} and \ref{ass:likelihood} hold, we have $\Vert \hat{V}_{n, N} - V \Vert_{\max} 
    \leq \mathcal{O}\left( n^{-1/2} \right) + \mathcal{O}\left( N^{-1/2} \right)$
almost surely.
\label{thm:ACD}
\end{theorem}
A proof of Theorem~\ref{thm:ACD} is provided in Appendix~\ref{app:sec:thmACD}. Provided the theorem holds, the two-stage approximation $\hat{V}_{n, N}$ will get closer to $V$ as the posterior sample size $n$ and the auxiliary sample size $N$ increase.

To provide theoretical justification for $\mathcal{\hat{C}}^{\textup{BM}}_{n, N}(\bx)$, we make the following assumption in addition to Conditions 1 and 2 of \citet{Vats2019}. 
\begin{assumption}
The batch size $b_{n,N}$ is an integer sequence that satisfies $b_{n,N} \to \infty$ and $(b_{n,N} / N)^{1/2}$ $\to$ $0$ as $N$ $\to$ $\infty$ and $n$ $\to$ $\infty$.
\label{ass:batchsize}
\end{assumption}
Under Assumptions~\ref{ass:prior}, \ref{ass:likelihood}, and \ref{ass:batchsize} and Conditions 1 and 2 of \citet{Vats2019},
Theorem~\ref{thm:ACD_bm} quantifies the distance between $\Sigma$ and $\hat{\Sigma}_{n,N}$. 
\begin{theorem}
Consider a sample $\bth^{(1)}, \dots, \bth^{(n)}$ generated from a Markov process that is an $\pi$-invariant polynomially ergodic and of order $m > (1+\epsilon_1)(1 + 2/\delta)$ for some $\epsilon_1 > 0$ and $\delta > 0$. 
Then Condition 1 of \citet{Vats2019} holds with $\gamma(n) = n^{1/2 - \lambda}$ for some $\lambda > 0$. If Assumptions~\ref{ass:prior}, \ref{ass:likelihood}, and \ref{ass:batchsize}, and Condition 2 of \citet{Vats2019} hold, we have
\begin{align*}
    \Vert \hat{\Sigma}_{n, N} - \Sigma \Vert_{\max} &\leq \mathcal{O}\left[  \{ (\log n) / b_{n,N} \}^{1/2} n^{1/2 - \lambda} \right] + \mathcal{O}\left[(b_{n,N}/N)^{1/2}\right]
\end{align*}
almost surely.
\label{thm:ACD_bm}
\end{theorem}
A proof of Theorem~\ref{thm:ACD_bm} is provided in Appendix~\ref{app:sec:thmACDbm}. Provided the theorem holds, the two-stage approximation $\hat{\Sigma}_{n, N}$ will get closer to $\Sigma$ as the posterior sample size $n$ and the auxiliary sample size $N$ increase.

A simple heuristic for determining $N$ is to plot the approximation $\hat{\bs{d}}_N(\bth)$ against the auxiliary sample size $N$ and select $N$ at which the approximation appears to stabilize. Similarly we choose $n$ at which the approximation $\hat{V}_{n,N}$ or $\hat{\Sigma}_{n,N}$ appears to stabilize for the pre-determined $N$. We henceforth take the batch size to be $b_{n,N} = \min\{n^{1/3}, N^{2/5}\}$ which satisfies Assumption~\ref{ass:batchsize} and Condition 2 of \citet{Vats2019}.

\section{A Kernel Stein Discrepancy}
\label{sec:KSD}
In this section we briefly describe an inverse multiquadric kernel Stein discrepancy (IMQ KSD) introduced by \citet{Gorham2017} which is a kernel Stein discrepancy-based diagnostic for assessing the convergence of a sequence of sample distributions to its target distribution and has theoretical support. To make this approach available for doubly intractable target distributions, we develop its Monte Carlo approximation, AIKS (approximate inverse multiquadric kernel Stein discrepancy) and show that it asymptotically inherits the same convergence properties as IMQ KSD.

\subsection{An Inverse Multiquadric Kernel Stein Discrepancy}
\label{subsec:IMQKSD}

Consider samples $\bth^{(1)}, \dots, \bth^{(n)}$ generated from an algorithm targeting a distribution $P$. Let $Q_n$ be the empirical distribution of the samples.
Suppose we want to evaluate $\textup{E}_P \left\{z(\bth)\right\}$, which is intractable. The weighted sample $Q_n$ provides an approximation $ \textup{E}_{Q_n}\left\{z(\bth) \right\} = \sum_{i=1}^n z(\bth^{(i)}) q_n(\bth^{(i)})$ of the target expectation. To assess the quality of the approximation, one may consider discrepancies quantifying the maximum expectation error over a set of test functions $\mathcal{Z}$: $d_{\mathcal{Z}}(Q_n, P) = \sup_{z \in \mathcal{Z}} \left| \textup{E}_{P} \left\{ z(\bth) \right\} - \textup{E}_{Q_n}\left\{ z(\bth) \right\} \right|$. When $\mathcal{Z}$ is large enough the discrepancy is called an integral probability metric (IPM) \citep{Muller1997} and $d_{\mathcal{Z}}(Q_n, P) \to 0$ only if $Q_n \Rightarrow P$ for any sequence $Q_n$. We use $\Rightarrow$ to denote the weak convergence of a sequence of probability measures. However, it is not practical to use IPM for assessing a sample since $\textup{E}_P\left\{ z(\bth) \right\}$ may not be computable for some $z  \in \mathcal{Z}$.

According to Stein's method \citep{Stein1972}, \citet{Gorham2015} defined a Stein discrepancy as $\mathcal{S}(Q_n, \mathcal{T}_P, \mathcal{G}) = \sup_{g \in \mathcal{G}} \left| \textup{E}_{Q_n} \left\{(\mathcal{T}_P g) (\bth) \right\} \right|$ for a Langevin Stein operator $\mathcal{T}_P$ and a Stein set $\mathcal{G}$ that satisfy $\textup{E}_P \left\{ (\mathcal{T}_P g) (\bth) \right\} = 0 $ for all $g \in \mathcal{G}$. The Stein discrepancy is the maximum expectation error over the Stein set $\mathcal{G}$ given the Stein operator $\mathcal{T}_P$. This avoids explicit integration under $P$ by selecting appropriate $\mathcal{T}_P$ and $\mathcal{G}$ that lead the target expectation to zero. \citet{Gorham2017} selected kernel Stein set based on the inverse multiquadric kernel and defined the inverse multiquadric kernel Stein discrepancy $\mathcal{S} (Q_n, \mathcal{T}_P, \mathcal{G}_{k, \Vert\cdot\Vert})$ for any norm $\Vert\cdot\Vert$, which admits closed-form solution.  
\begin{definition}[IMQ KSD \citep{Gorham2017}] 
Let $k(\bx, \by) = (c^2 + \Vert\bx - \by \Vert_2^2 )^\beta$ for some $c > 0$ and $\beta \in (-1,0)$.
For each $j \in \{1, \dots, p\}$ construct the Stein kernel
\begin{align*}
    k_0^j(\bx, \by) &= u_j(\bx) u_j(\by) k(\bx, \by) + u_j(\bx) \nabla_{y_j} k(\bx, \by) + u_j(\by) \nabla_{x_j} k(\bx, \by) + \nabla_{x_j} \nabla_{y_j} k(\bx, \by),
\end{align*}
where $u_j$ is the $j$th entry of the score function of the target density. Then IMQ KSD $\mathcal{S}(Q_n, \mathcal{T}_P, \mathcal{G}_{k, \Vert\cdot\Vert} ) = \Vert\textbf{w}\Vert$, where $\text{w}_j = \sqrt{ \frac{1}{n^2} \sum_{k,l = 1}^n k_0^j(\bth^{(k)}, \bth^{(l)})}$. 
\end{definition}
Computation of $\textbf{w}$ is parallelizable over sample pairs $(\bth^{(k)}, \bth^{(l)})$ and coordinates $j$. \citet{Gorham2017} provided theoretical justification for its use for diagnosing convergence of a sequence $Q_n$ to its target distribution $P$ (see Theorem 8 and Proposition 9 of \citet{Gorham2017}).
\subsection{An Approximate Inverse Multiquadric Kernel Stein Discrepancy for Intractable Normalizing Function Problems}
\label{subsec:AIKS}

When the target distribution $P$ is doubly intractable, computation of IMQ KSD is not feasible. This is because IMQ KSD requires evaluating the score function of the target density. In this section we introduce an approximate version of IMQ KSD with $L^2$ norm by replacing the score function with its Monte Carlo approximation. The approximate inverse multiquadric kernel Stein discrepancy (AIKS) is defined as follows.
\begin{definition}[AIKS]
Let $k(\bx, \by)$ $=$ $(c^2 + \Vert\bx - \by \Vert_2^2 )^\beta$ for some $c > 0$ and $\beta \in (-1,0)$. Define an approximate Stein kernel as
    \begin{align*}
        \hat{k}_0(\bx, \by) &= \sum_{j=1}^p \bigg\{ \hat{u}_j(\bx) \hat{u}_j(\by) k(\bx, \by) + \hat{u}_j(\bx) \nabla_{y_j} k(\bx, \by) + \hat{u}_j(\by) \nabla_{x_j} k(\bx, \by) + \nabla_{x_j} \nabla_{y_j} k(\bx, \by) \bigg\},
    \end{align*}
    where $\hat{u}_j$ is the $j$th entry of the approximate score function of the target density. 
    Then the AIKS is defined as $\hat{\mathcal{S}}(Q_n, \mathcal{T}_P, \mathcal{G}_{k, \Vert\cdot\Vert_2} ) = \frac{1}{n^2} \sum_{k,l=1}^{n} \hat{k}_0(\bth^{(k)}, \bth^{(l)}).$
\end{definition}

For doubly intractable target density $\pi(\bth \mid \bx)$, the score function can be approximated as described in Section~\ref{subsec:ACD}. Under the assumptions of IMQ KSD on the target distribution, Theorem~\ref{thm:AIKS} quantifies the distance between AIKS and IMQ KSD for $L^2$ norm. 
\begin{theorem}
For a target distribution $P$ having a bounded score function and a sample distribution $Q_n$ targeting $P$, let $\mathcal{S}(Q_n, \mathcal{T}_P, \mathcal{G}_{k, \Vert\cdot\Vert_2} )$ and $\hat{\mathcal{S}}(Q_n, \mathcal{T}_P, \mathcal{G}_{k, \Vert\cdot\Vert_2} )$ denote IMQ KSD and AIKS, respectively, for the sample distribution. Then
\begin{align*}
    \left| \hat{\mathcal{S}}(Q_n, \mathcal{T}_P, \mathcal{G}_{k, \Vert\cdot\Vert_p} ) - \mathcal{S}(Q_n, \mathcal{T}_P, \mathcal{G}_{k, \Vert\cdot\Vert_p} ) \right| &\leq \mathcal{O}\left\{ N^{-1/4} \right\}  
\end{align*}
almost surely where $N$ is the auxiliary sample size used to approximate the score function of the target density.
\label{thm:AIKS}
\end{theorem}
A proof of Theorem~\ref{thm:AIKS} is provided in Appendix~\ref{app::sec:thmAIKS}. The theorem implies that AIKS will get closer to IMQ KSD as the auxiliary sample size $N$ increases. 
The following proposition shows that AIKS asymptotically inherits the same convergence properties as IMQ KSD.
\begin{proposition}
    \begin{enumerate}
    \item[(i)] Detecting non-convergence: for a distantly dissipative target distribution $P$, if $\hat{\mathcal{S}}(Q_n, \mathcal{T}_P, \mathcal{G}_{k, \Vert\cdot\Vert_2} )$ $\to$ $0$ as $n,N \to \infty$, then $Q_n \Rightarrow P$ as $n \to \infty$.
    \item[(ii)] Detecting convergence: for a target distribution $P$ having Lipschitz score function with $\textup{E}_P \left\{ \Vert\bs{u}(\bth)\Vert_2^2 \right\} < \infty$, if the Wasserstein distance $d_{\mathcal{W}_{\Vert\cdot\Vert_2}}(Q_n, P) \to 0$ as $n \to \infty$, then $\hat{\mathcal{S}}(Q_n, \mathcal{T}_P, \mathcal{G}_{k, \Vert\cdot\Vert_2} ) \to 0$ as $n,N \to \infty$.
\end{enumerate}
\label{prop:AIKS}
\end{proposition}
A proof of Theorem~\ref{prop:AIKS} is provided in Appendix~\ref{app::sec:propAIKS}. In this article we use $\beta = -1/2$, and $c = 1$.

The AIKS is a form of two-sample $V$-statistics. \citet{Chwialkowski2016} obtain the asymptotic distribution of a kernel Stein discrepancy using the theory of $V$-statistics for dependent samples. \citet{Leucht2013} provides the asymptotic distribution of a $V$-statistic for dependent samples. By a direct application of Theorem 2.1 of \citet{Leucht2013}, the following proposition characterizes the asymptotic behavior of AIKS.
\begin{proposition}
    Consider a sample $\bth^{(1)}, \dots, \bth^{(n)}$ generated from a $P$-invariant $\beta$-mixing process with $\sum_{r=1}^\infty r^2 \sqrt{\beta(r)} < \infty$. If $\hat{k}_0$ is Lipschitz continuous and $\text{E}_P[\hat{k}_0(\bth, \bth)] < \infty$, then as $n,N \to \infty$, we have $n \hat{\mathcal{S}}( Q_{n}, \mathcal{T}_P, \mathcal{G}_{k,\Vert \cdot \Vert_2}) \stackrel{d}{\to} \sum_{k=1}^\infty \lambda_k Z^2_k$ almost surely. The $\{\lambda_k\}$ are the eigenvalues of kernel $\hat{k}_0(\bth, \bth^{\prime})$ under P, i.e., they are the  solutions of $\lambda_k \phi_k(\bth) = \int_{\bth_r} \hat{k}_0(\bth, \bth_r) \phi_k(\bth_r) P(\bth_r) d\bth_r$. The $\{Z_k\}$ are centered, jointly Gaussian random variables with $\text{cov}(Z_i,Z_j) = \sum_{r=-\infty}^{\infty} \text{cov} [ \phi_i(\bth), \phi_j(\bth_r) ]$.
\label{prop:asyDistAIKS}
\end{proposition}
$\beta$-mixing is a notion of dependence that is weak enough for most practical uses. For instance, iid samples and stationary, geometrically ergodic Markov chains are $\beta$-mixing. The proof of Proposition~\ref{prop:asyDistAIKS} is a simple verification of assumptions and can be found in the Appendix~\ref{app:sec:prop:asyDistAIKS}.

However, the asymptotic distribution of AIKS does not have a closed form in general. To obtain consistent estimates of quantiles of the asymptotic distribution, we employ a bootstrap method proposed by \citet{Leucht2013}. We compute bootstrap samples as $\hat{\mathcal{S}}^{\ast} = \frac{1}{n^2} \sum_{k,l=1}^{n} (W_k-\bar{W}) \hat{k}_0(\bth^{(k)}, \bth^{(l)}) (W_l-\bar{W})$, where $\bar{W} = \frac{1}{n}\sum_{k=1}^{n} W_k$ and $\{W_k, k = 1,\dots,n\}$ is an auxiliary random process for accommodating the dependence structure of the original process $\{ \bth^{(k)},k=1,\dots,n \}$ asymptotically when the kernel $\hat{k}_0$ is degenerate, i.e., when the sample is drawn from a $P$-invariant process. By Remark 2 of \citet{Leucht2013}, we can construct the auxiliary process as $W_k = e^{-1/\xi} W_{k-1} + \sqrt{ 1 - e^{-2/\xi} } \epsilon_k$ where $W_0,\epsilon_1,\dots,\epsilon_n$ are independent standard normal random variables and $\xi$ is a tuning parameter that performs a similar role to the block length of blockwise bootstrap methods. A simulation experiment by \citet{Leucht2013} shows that the choice of $\xi$ does not significantly affect test performance. We use $\xi = 7$ in the sequel. We calculate the empirical $(1-\alpha)$ quantile $\hat{\gamma}_{1-\alpha}$ of $n \hat{\mathcal{S}}^\ast$. We use $\hat{\gamma}_{1-\alpha}$ as a threshold for AIKS. The following proposition shows that $\hat{\gamma}_{1-\alpha}$ consistently approximates the quantile of the limiting distribution of $n \hat{\mathcal{S}}( Q_{n}, \mathcal{T}_P, \mathcal{G}_{k,\Vert \cdot \Vert_2})$.
\begin{proposition}
    Consider a sample $\bth^{(1)}, \dots, \bth^{(n)}$ generated from a $P$-invariant $\beta$-mixing process with $\sum_{r=1}^\infty r^2 \sqrt{\beta(r)} < \infty$. 
    If $\hat{k}_0$ is Lipschitz continuous and $\text{E}_P[\hat{k}_0(\bth, \bth)] < \infty$, then as $n,N \to \infty$, we have $\sup_x \big\vert \text{P}(n \hat{\mathcal{S}}^{\ast} \leq x ) - \text{P}(n \hat{\mathcal{S}}(Q_n, \mathcal{T}_P, \mathcal{G}_{k,\Vert \cdot \Vert_2}) \leq x ) \big \vert \stackrel{p}{\to} 0$.
    \label{prop:bootAIKS}
\end{proposition}
The proof is a straightforward verification of assumptions that can be found in the Appendix~\ref{app:sec:prop:bootAIKS}. A sample path for which $n \hat{\mathcal{S}}( Q_{n}, \mathcal{T}_P, \mathcal{G}_{k,\Vert \cdot \Vert_2})$ is below $\hat{\gamma}_{1-\alpha}$ is considered to have an asymptotic distribution that is reasonably close to the target distribution.

A notable special case of doubly intractable posterior distributions arises when a prior distribution has an intractable normalizing constant that is a function of parameter of interest \citep[cf.][]{Rao2016,Vats2022}. Suppose that we have a prior density $p(\bth \mid \bs{\eta}) = g(\bth \mid \bs{\eta}) / z(\bs{\eta})$, where $z(\bs{\eta})$ is intractable. If hyperprior $k(\cdot)$ is assigned to $\bs{\eta}$, the posterior density is
\begin{align}
    \pi(\bth, \bs{\eta} \mid \bx) &\propto \frac{g(\bth \mid \bs{\eta})}{z(\bs{\eta})} k(\bs{\eta}) L(\bth \mid \bx), \nonumber
\end{align}
which leads to a doubly intractable posterior distribution. Our diagnostics ACD and AIKS can be applied to this case if auxiliary variables can be generated exactly from the prior distribution or generated by a Monte Carlo algorithm having $p(\bth \mid \bs{\eta})$ as its target distribution. Using the auxiliary variables, we can approximate intractable derivatives of the prior and obtain our diagnostic quantities.

\section{Comparison of Diagnostics}
\label{sec:comp}

\begin{table}[!t]
\centering
\caption{Power and Type I error rates of a multivariate normality test with increasing sample size $n$ and dimension $p$, averaged over 100 simulations.}
\label{tab:powerNerror}
\begin{tabular}{lc rrrrrr c rrrrrr}
    \toprule
    \multirow{2}{*}{Method} & \multirow{2}{*}{$n$} & \multicolumn{6}{c}{Power} && \multicolumn{6}{c}{Type I error rate} \\ 
    \cmidrule{3-8} \cmidrule{10-15}
    & & $p$=2 & 5 & 10 & 15 & 20 & 25 && 2 & 5 & 10 & 15 & 20 & 25 
    \\\midrule
    \multirow{3}{*}{CD}  & 1000 & 1 & 1 & 1 & 1 & 1 & 1 && 0.01 & 0.05 & 0.04 & 0.07 & 0.09 & 0.13 \\
    & 2000 & 1 & 1 & 1 & 1 & 1 & 1 && 0.00 & 0.00 & 0.01 & 0.05 & 0.04 & 0.10 \\ 
    & 5000 & 1 & 1 & 1 & 1 & 1 & 1 && 0.01 & 0.01 & 0.00 & 0.01 & 0.00 & 0.02 
    \\\midrule
    \multirow{3}{*}{KSD}  & 1000 & 1 & 1 & 1 & 1 & 1 & 1 && 0.00 & 0.02 & 0.00 & 0.00 & 0.00 & 0.00 \\ 
    & 2000 & 1 & 1 & 1 & 1 & 1 & 1 && 0.00 & 0.03 & 0.00 & 0.00 & 0.00 & 0.00 \\ 
    & 5000 & 1 & 1 & 1 & 1 & 1 & 1 && 0.01 & 0.05 & 0.00 & 0.00 & 0.00 & 0.00 \\ 
    \bottomrule
\end{tabular}
\end{table}

We recreate the experiment in Section 4 of \citet{Chwialkowski2016} and compare the power and Type I error rates of the CD-based and IMQ KSD-based tests. For $i = 1,\dots,n$ we generate $\bth^{(i)} = \bs{z}^{(i)} + u_i \bs{e}_1$, where $\bs{z}^{(i)} \stackrel{\text{iid}}{\sim} \text{MVN}(\textbf{0}, I_{p})$, $u_i \stackrel{\text{iid}}{\sim} \text{Unif}(0,1)$, and $\bs{e}_1 = (1,0,\dots,0)^\top$, for various combinations of sample size $n$ and parameter dimension $p$. We compare the power of the CD test and IMQ KSD test to assess whether $\bth^{(1)},\dots,\bth^{(n)}$ were drawn from the target distribution $P = \text{MVN}(\textbf{0}, I_p)$. We also generate samples from $P$ and compare the Type I error rates of the two tests. The nominal significance level is $\alpha = 0.01$. The results, averaged over 100 simulations, are displayed in Table~\ref{tab:powerNerror}. Both tests provide high power for the values of $n$ and $p$ we considered. We see that the Type I error rate of the CD grows as $p$ increases and $n \leq 2000$.  The CD requires approximation of the asymptotic covariance matrix of $\bs{d}(\bth)$, which is a $p(p+1)/2$ by $p(p+1)/2$ matrix. A sufficiently large $n$ is required to accurately estimate this covariance matrix. The Type I error rate of the CD is well controlled when $n = 5000$. We see that the IMQ KSD's Type I error rate is well controlled across the range of $n$ and $p$ values. We recommend AIKS when $n$ is small but $p$ is high.

We note that violation of higher-order identities, corresponding to the third moment or higher, cannot be detected by our curvature diagnostic because our diagnostic does not verify that any moment constraint of order 3 or higher is satisfied. Extending our methodology to include higher-order Bartlett identities is left to future work. The AIKS is supported by a theory of weak convergence and thus can be used for deciding convergence of a sequence of sample points to the target distribution.

A limitation of the ACD is that it requires second-order gradients of the likelihood and prior while AIKS requires only the first-order gradient.

The ACD is more useful than AIKS in some cases. When obtaining a gold standard sample is very expensive or impossible, it is difficult to obtain a good threshold for AIKS. If samples were generated from the target distribution, the ACD is asymptotically $\chi^2(r)$ distributed, where $r = p(p+1)/2$ and $p$ is the parameter dimension. The threshold value is the $(1-\alpha)$ quantile of the $\chi^2(r)$ distribution, which depends only on the parameter dimension $p$ and significance level $\alpha$. However, the asymptotic distribution of AIKS has no closed-form expression. We use a bootstrap procedure to approximate the $(1-\alpha)$ quantile of the limiting distribution. The approximation is consistent when the samples used for the bootstrap were drawn from the target distribution. Empirically, this approach performs fairly well if we use samples generated from a distribution that is reasonably close to the target distribution. 
(In our examples, we used samples generated from the exchange algorithm (which is asymptotically exact) or DMH (with $m$ large enough) to obtain the threshold.) If sample quality is low, however, we cannot obtain a suitable threshold.

The ACD is more computationally efficient than AIKS for large $n$. Let $s$ denote the computational cost of simulating a single auxiliary variable. Using $N$ auxiliary variables, we approximate $\bs{d}(\bth)$ for ACD and $\bs{u}(\bth)$ for AIKS at each posterior sample point. The time complexity of the approximation step is $\mathcal{O}(nsN)$. The time complexity for ACD is $\mathcal{O}(n)$, and the time complexity for AIKS is $\mathcal{O}(n^2)$, given the approximations.

\section{Applications}
\label{sec:app}
Here we apply our methods to both asymptotically exact and asymptotically inexact algorithms in the context of three general classes of models with intractable normalizing functions: (1) the Ising model, (2) a social network model, and (3) a Conway--Maxwell--Poisson regression model. Effective sample size (ESS), which is one of the most widely used MCMC diagnostics, is helpful for asymptotically exact methods since all chains converge to the target distribution. However, for asymptotically inexact methods, ESS is inadequate since an algorithm that mixes better might yield a poorer approximation to the target distribution. To illustrate the usefulness of our approaches for asymptotically inexact methods, we compare the approximate curvature diagnostic (ACD) and the approximate multiquadric kernel Stein discrepancy (AIKS) with ESS. 
We also study the computational complexity and the variability of our diagnostics. The code for our diagnostics is implemented in {\tt R} \citep{Ihaka1996} and {\tt C++}, using the {\tt Rcpp} and {\tt RcppArmadillo} packages \citep{Eddelbuettel2011}. The calculation of ESS follows \citet{kass1998} and \citet{Robert2004}. All code was run on dual 10-core Xeon E5-2680 processors on the Penn State high-performance computing cluster. 
The source code may be found in the following repository (\texttt{https://github.com/bokgyeong/Diagnostics}).

\subsection{The Ising Model}
\label{subsec:ising}

The Ising model \citep{lenz1920, ising1925} is one of the most famous and important models from statistical physics and provides an approach for modeling binary images. For an $r\times s$ lattice $\bx = \{x_{ij}\}$ with binary values $x_{i,j} \in \{-1, 1\}$, where $i, j$ denotes the row and column, the Ising model with a parameter $\theta \in (0,1)$ has likelihood 
\[
    L(\theta \mid \bx) = \frac{1}{c(\theta)}\exp\{\theta S(\bx)\},
\]
where
\[
    S(\bx) = \sum_{i=1}^r \sum_{j=1}^{s-1} x_{i,j} x_{i,j+1} + \sum_{i=1}^{r-1} \sum_{j=1}^{s} x_{i,j} x_{i+1,j}, 
\]
is the sum over all possible products of neighboring elements and imposes spatial dependence. A larger value for the dependence parameter $\theta$ produces stronger interactions among neighboring observations. Calculation of the normalising function $c(\theta)$ requires summation over all $2^{rs}$ possible outcomes for the model, which is computationally infeasible even for lattices of moderate size. We carried out our simulation using perfect sampling \citep{propp1996} on a $30 \times 30$ lattice with the parameter setting $\theta=0.2$, which represents moderate dependence.

\begin{figure}[!tb]
\centering
\includegraphics[width=\textwidth]{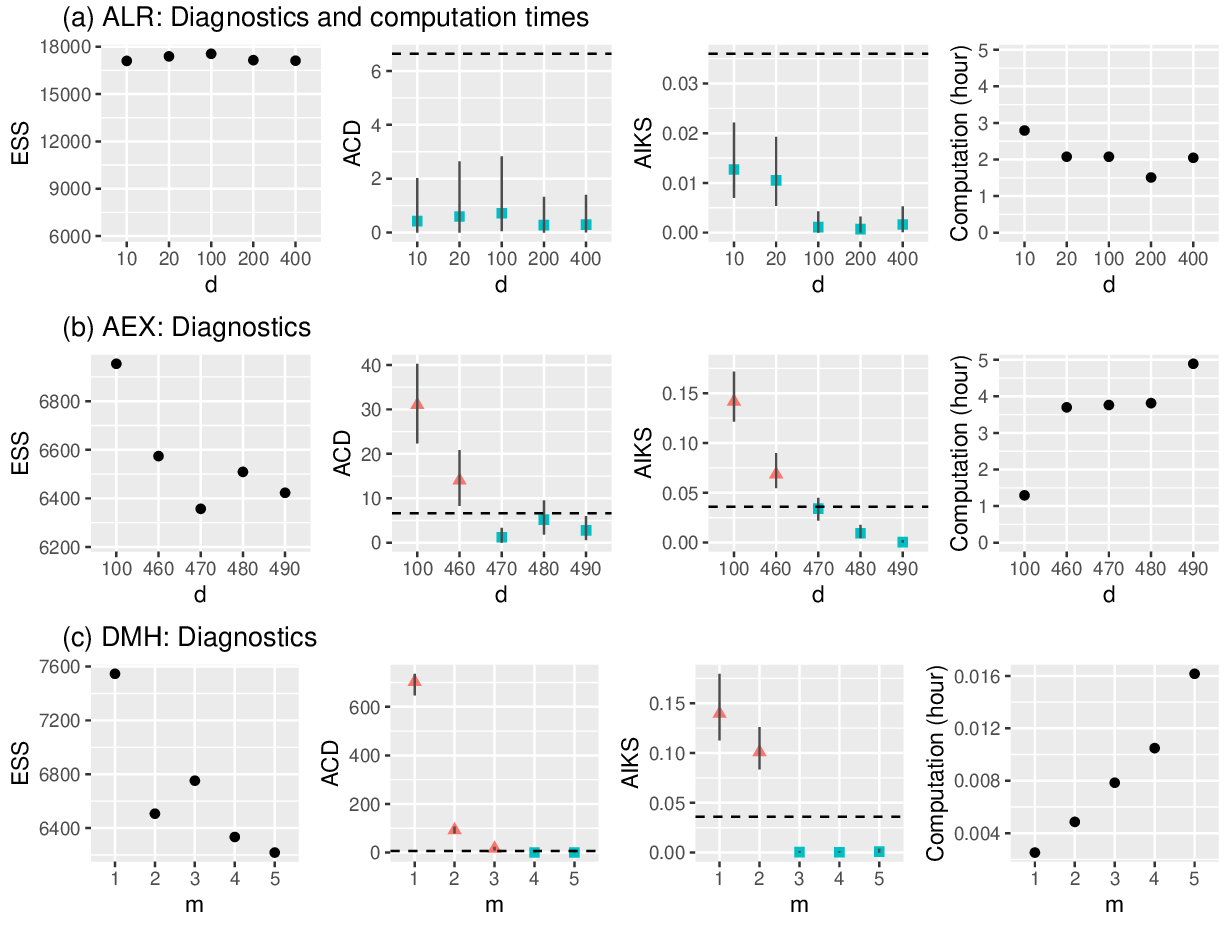}
\caption{Results for the Ising model. The ESS, ACD, AIKS, and computation time for samples generated from (a) ALR with different numbers $d$ of particles, (b) AEX with different numbers $d$ of particles, and (c) DMH with different numbers $m$ of inner updates. For ACD and AIKS, the dashed horizontal line represents the threshold value. The triangle/square and vertical line show the empirical mean and 95\% uncertainty interval, respectively, of 30 replications of the diagnostic. The red triangle and blue square indicate poor sample quality and good sample quality, respectively. 
}
\label{fig:ising}
\end{figure}

For this example we consider ALR, AEX, and DMH algorithms described in Section~\ref{sec:intractable normalizing functions}. The ALR and AEX algorithms are asymptotically exact. The DMH algorithm is asymptotically inexact but was found to be very efficient in terms of effective sample size per unit time and is applicable for doubly intractable problems with higher parameter dimension. To find appropriate values for the tuning parameters of the algorithms, we generate multiple chains from each algorithm with different choices of tuning parameters. 
The performance of ALR and AEX depends heavily on the set of particles. A sufficiently large number $d$ of particles is required to cover the important region of the parameter space. Selecting a suitable value of $d$ is difficult because said value varies across models or across datasets for a given model. The choices for the other tuning parameters are rather well studied and good heuristics exist. We implement ALR and AEX with different numbers $d$ of particles. We choose values for the other tuning parameters according to \citet{Park2018}.
To obtain the particles, we used fractional DMH, which is DMH with a larger acceptance probability, and a short run of DMH with a single cycle of (inner) Gibbs updates for AEX and ALR, respectively.
We implement DMH with different numbers $m$ of (inner) Gibbs updates. We use a uniform prior with range $[0,1]$ for $\theta$. 

All algorithms were run for $n$ = 100,000 iterations, and all 100,000 posterior samples were used for testing.
We used parallel computing to obtain 30 replications of each diagnostic for each sample path. We assess the samples using the empirical mean of 30 replications. The threshold value of ACD is the 0.99 quantile of $\chi^2(1)$, which is 6.63. We use the sample generated from the exchange algorithm \citep{murray2006}, which is asymptotically exact and has no tuning parameter, as the gold standard. Using the sample, we estimated the 0.99 quantile of the asymptotic distribution of AIKS via the bootstrap method, which provided 0.036. ACD computation took approximately 3 seconds and AIKS computation took approximately 27 seconds.

\setlength{\tabcolsep}{4.6pt}
\begin{table}[!tb]
    \centering
    \caption{Summary statistics of posterior samples and a gold standard in the Ising model for a simulated binary lattice.}
    \label{tab:ising}
    \begin{tabular}{lrrrr}
        \toprule
        Algorithm & $d$ or $m$ & Median & LTP & RTP\\\midrule
        DMH & 1 & 0.19 & 0.09 & 0.09\\ 
        DMH & 4 & 0.19 & 0.05 & 0.05\\
        ALR & 10 & 0.19 & 0.05 & 0.05 \\
        AEX & 470 & 0.19 & 0.05 & 0.04 \\
        \midrule
        Gold standard & & 0.19 & 0.05 & 0.05 \\\bottomrule
    \end{tabular}

    \smallskip
    {\raggedright SD, standard deviation; LTP, left-tail probability; RTP, right-tail probability. \par}
\end{table}

Figure~\ref{fig:ising} shows the diagnostic values and computation time for each sample path. For ALR, Figure~\ref{fig:ising} (a) shows that all diagnostics provide the same conclusion: ALR is not so sensitive to the number $d$ of particles in this setting. For AEX and DMH, ACD and AIKS provide considerably different conclusions from ESS. Figure~\ref{fig:ising} (b) we see that, for AEX, ESS is maximized (the best) at $d$ = 100 and generally decreases (worsens) as $d$ increases. On the other hand, ACD and AIKS have their largest (worst) values at $d =$ 100 and indicate that $d$ should be at least 470. Likewise, for DMH, Figure~\ref{fig:ising} (c) shows that ESS is maximized (the best) at $m$ = 1 and generally decreases (worsens) as $m$ increases, while ACD and AIKS take their largest (worst) values at $m =$ 1 and decrease (improve) as $m$ increases. ACD recommends $m$ of 4 or more, and AIKS recommends $m$ of 3 or more. 
The computation time is measured using a single run for each case. We observe that DMH with $m = 4$ provides a good approximation of the target distribution while being extremely computationally efficient.

Table~\ref{tab:ising} presents summary statistics for some of the posterior samples and the gold standard. Cutoff values for the left- and right-tail probabilities are the lower 5\% and the upper 5\% percentiles of the gold standard. Both ACD and AIKS suggest that ALR with $d =$ 10, AEX with $d = $ 470, and DMH with $m =$ 4 provide high quality samples. It is observed that they provide similar values of the summary statistics to the gold standard. 
On the other hand, the ESS-recommended DMH sample with $m =$ 1 does not match the gold standard. It provides higher tail probabilities compared to the gold standard. This shows that ACD and AIKS perform well in assessing how close samples are to the exact posterior. Not surprisingly, ESS is inadequate as a tool for this purpose. In summary, our approaches provide good guidance on how to assess the quality of samples for both the asymptotically exact and asymptotically inexact algorithms. In addition, our diagnostics help one to select the best algorithm and its tuning parameter in terms of statistical efficiency.

\subsection{A Social Network Model}
\label{subsec:ergm}

An exponential family random graph model (ERGM) \citep{Robins2007, Hunter2008} is a statistical model for analysing network data. Consider an undirected ERGM with $n$ vertices. Relationships among the vertices can be represented as an $n\times n$ adjacency matrix $\bx$ as follows: for all $i \neq j$, $x_{i,j} = 1$ if the $i$th and  $j$th vertices are connected, and $x_{i,j} = 0$ otherwise. And $x_{i,i}$ is 0 for all $i \in \{1,\dots,n\}$, i.e., there are no loops. The number of possible network configurations is $2^{n(n-1)/2}$ and summation over those configurations is required to calculate the normalizing function of the model. Thus computing $c(\bth)$ is infeasible unless $n$ is quite small. 

For an undirected graph with $n$ vertices, the ERGM likelihood is given by
\begin{align*}
    L(\bth \mid \bx) &= \frac{1}{c(\bth)}\exp\{\theta_1 S_1(\bx) + \theta_2 S_2(\bx)\},\\
    S_1(\bx) &= \sum_{i=1}^n \binom{x_{i+}}{1},\\
    S_2(\bx) &= e^\tau \sum_{k=1}^{n-2} \left\{ 1 - (1 - e^{-\tau})^k \right\} \textup{EP}_k (\bx).
\end{align*}
The sufficient statistics $S_1(\bx)$ and $S_2(\bx)$ are the number of edges in the graph and the geometrically weighted edge-wise shared partnership (GWESP) statistic \citep{Hunter2006,Hunter2007}, respectively. $x_{i+}$ denotes the sum of the $i$th row of the adjacency matrix and $\textup{EP}_k(\bx)$ denotes the number of edges between two vertices that share exactly $k$ neighbors. It is assumed that $\tau$ is fixed at a value of 0.25. We simulated a network with 30 actors using 10,000 cycles of Gibbs updates, where the true parameter was $\bth=(\theta_1, \theta_2)^\top = (-0.96, 0.04)^\top$.

\begin{figure}[!t]
\centering
\includegraphics[width = \textwidth]{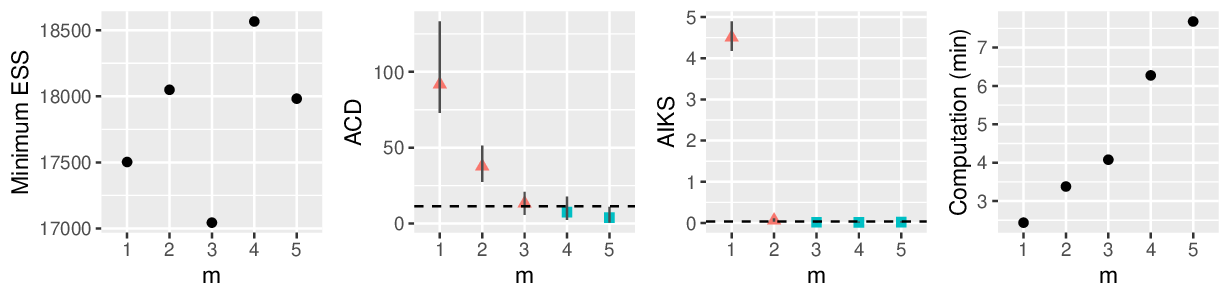}
\caption{
Results for the ERGM. The minimum ESS, ACD, AIKS and computation time for samples generated from DMH with different numbers $m$ of inner updates. For ACD and AIKS, the dashed horizontal line represents the threshold value. The triangle/square and vertical line show the empirical mean and 95\% uncertainty interval, respectively, of 30 replications of the diagnostic. The red triangle and blue square indicate poor sample quality and good sample quality, respectively.
}
\label{fig:ergm}
\end{figure}

For this example we consider the DMH algorithm, which is asymptotically inexact. We explained the DMH algorithm in Section~\ref{subsec:ising}. We implement DMH with different numbers $m$ of Gibbs updates. We use uniform priors with range $[-5.00, 2.27] \times [-1.57, 2.32]$ for $(\theta_1, \theta_2)^\top$, which are centered around the maximum pseudo-likelihood estimates (MPLE) and have widths of 12 standard deviations. 

All algorithms were run for $n$ = 300,000 iterations, and all 300,000 posterior samples were used for testing.
We used parallel computing to obtain 30 replications of each diagnostic for each sample path. We assess the samples using the empirical mean of 30 replications. The threshold value of ACD is the 0.99 quantile of $\chi^2(3)$, which is 11.34. We treat a run from DMH with 20 cycles of Gibbs updates as the gold standard.
Using the sample, we estimated the 0.99 quantile of the asymptotic distribution of AIKS via the bootstrap method, which provided 0.035. ACD computation took approximately 54 seconds and AIKS computation took approximately 38 minutes.

Figure~\ref{fig:ergm} shows the minimum effective sample size---this is the conservative way of using ESS when there are multiple ESSs due to the multivariate posterior distribution---, ACD, AIKS, and computation time for the DMH sample, for a sequence of $m$ values. The minimum ESS is maximized (the best) at $m =$ 4. ACD and AIKS generally decrease (become better) as $m$ increases. The ACD implies that $m$ should be at least 4, and AIKS suggests that $m$ should be at least 3. 
We see that $m = 4$ yields a good approximation according to both diagnostics, at a modest computing cost.
We observe that all ESS-, ACD-, and AIKS-selected samples provide almost the same values of the summary statistics as the gold standard. In this case, all of the diagnostics generally agree.

\subsection{A Conway--Maxwell--Poisson Regression Model}
\label{subsec:comp}

The Conway--Maxwell--Poisson (COMP) distribution \citep{Conway1962} is a two-parameter generalisation of the Poisson distribution for modeling count data that are characterised by under-dispersion (variance less than the mean) or over-dispersion (variance greater than the mean). For a COMP($\lambda$, $\nu$) variable $Y$, the probability mass function is given by
\begin{align}
    P(Y = y ) &= \frac{1}{c(\lambda, \nu)} \frac{\lambda^y}{(y!)^{\nu}} \nonumber,
\end{align}
where $\lambda > 0$ is a generalisation of the Poisson rate parameter, $\nu \geq 0$ denotes the dispersion parameter, and $c(\lambda, \nu) = \sum_{z=0}^{\infty} \lambda^z / (z!)^{\nu}$ is a normalising function. The COMP distribution accommodates under- ($\nu >$ 1), equi- ($\nu =$ 1), or over-dispersion (0 $\leq \nu <$ 1). The COMP distribution forms a continuous bridge between the  Poisson ($\nu =$ 1), geometric ($\nu =$ 0 and $\lambda <$ 1), and Bernoulli ($\nu =$ $\infty$ and success probability $\lambda/(1+\lambda)$) distributions. \citet{Guikema2008} proposed a reparameterisation, substituting $\eta$ = $\lambda^{1/\nu}$ to approximate the center of the COMP distribution. For a count variable $Y$ that follows the COMP$_{\eta}$($\eta$, $\nu$) distribution, the probability mass function is
\begin{align}
    P(Y = y) &= \frac{1}{c_{\eta}(\eta, \nu)} \left(\frac{\eta^y}{y!}\right)^{\nu} \nonumber,
\end{align}
where $c_{\eta}(\eta, \nu) = \sum_{z=0}^{\infty} \left(\eta^z/ z! \right)^{\nu}$ is the normalising function. \citet{Huang2017} and \citet{Ribeiro2020} have reparameterised the COMP distribution as a function of the mean. Under any parameterisation, however, the COMP normalising function is an infinite sum, making the function intractable.

For count response variables $Y_i$ and corresponding explanatory variables $\bx_i$ $=$ ($x_{i,1}$, $\dots$, $x_{i,p-1}$)$^\top$ for $i$ $=$ $1$, $\dots$, $n$, the likelihood of a COMP regression model with log link function for $\eta$ is given by $Y_i \sim \text{COMP}_{\eta} (\eta_i, \nu)$, with $\log(\eta_i) = \beta_0 + x_{i,1}\beta_1 + \cdots + x_{i,p-1}\beta_{p-1} \; (i = 1,\dots,n)$, where $\bs{\beta}$ $=$ ($\beta_0$, $\dots$, $\beta_{p-1}$)$^\top$ are regression coefficients. We study the takeover bids dataset \citep{Cameron1997}, which comprises the number of bids received by 126 US firms that were targets of tender offers during the period 1978--1985, along with 9 explanatory variables \citep[see][for details]{Saez-Castillo2013}. The dataset can be obtained from \texttt{mpcmp} package \citep{fung2019mpcmp}. It is assumed that $\nu$ is fixed at a value of 1.754 \citep{Huang2017}.

\begin{figure}[!t]
\centering
\includegraphics[width=\textwidth]{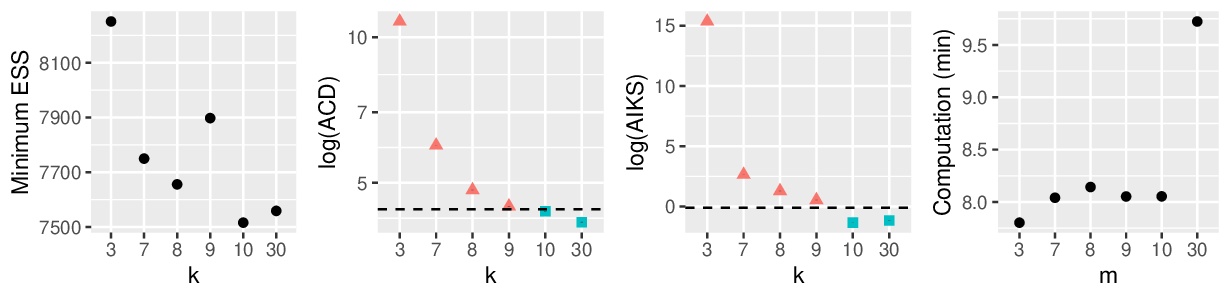}
\caption{Results for the COMP regression model. The minimum ESS, ACD, AIKS and computation time for samples generated from NormTrunc with different levels $k$ of truncation. For ACD and AIKS, the dashed horizontal line represents the threshold value. The triangle/square and vertical line show the empirical mean and 95\% uncertainty interval, respectively, of 30 replications of the diagnostic. The red triangle and blue square indicate poor sample quality and good sample quality, respectively.
}
\label{fig:comp}
\end{figure}

For this example we consider a simple and widely used approach, namely, the NormTrunc algorithm. Specifically, we approximate the normalizing function by truncating the infinite sum to a truncation level $k$ such that $c_{\eta}(\eta, \nu)$ $\approx$ $\sum_{z=0}^k \left( \eta^z / z! \right)^{\nu}$ and use the approximation for each Metropolis--Hastings accept-reject ratio. The NormTrunc algorithm is asymptotically inexact. In order to determine a suitable value of $k$, we implement NormTrunc with different levels $k$ of truncation.

All algorithms were run for $n$ = 300,000 iterations, and all 300,000 posterior samples were used for testing.
We used parallel computing to obtain 30 replications of each diagnostic for each sample path. We assess the samples using the empirical mean of 30 replications. The threshold value of ACD is the 0.99 quantile of $\chi^2(55)$, which is 82.29. We treat as the gold standard a run from the exchange algorithm \citep{murray2006}. Using the sample, we estimated the 0.99 quantile of the asymptotic distribution of AIKS via the bootstrap method, which provided 0.9. ACD computation took approximately 5.8 hours and AIKS computation took approximately 6.3 hours.

Figure~\ref{fig:comp} shows minimum effective sample size, ACD, AIKS, and computing time for the NormTrunc sample for a range of $k$ values. The minimum ESS is maximized (the best) at $k =$ 3. On the other hand, both ACD and AIKS recommend $k$ of 10 or more. 
We see that the uncertainty of our diagnostics is very small since we generate auxiliary variables exactly from the model using the rejection sampler. It appears that $k = 10$ is sufficient.

\begin{table}[!t]
    \centering
    \caption{Summary statistics of posterior samples and a gold standard for $\beta_0$ in the COMP regression model.}
    \label{tab:comp}
    \begin{tabular}{lrrrr}
        \toprule
        Algorithm & $k$ & Median & LTP & RTP\\\midrule
        NormTrunc & 3 & 0.45 & 0.01 & 0.47\\
        NormTrunc & 10 & 0.31 & 0.05 & 0.05\\\midrule
        Gold standard & & 0.31 & 0.05 & 0.05 \\\bottomrule
    \end{tabular}

    \smallskip
    {\raggedright SD, standard deviation; LTP, left-tail probability; RTP, right-tail probability. \par}
\end{table}

Table~\ref{tab:comp} presents summary statistics for some of the posterior samples and a gold standard. The cutoff values for left- and right-tail probabilities were chosen as in the previous example. Both ACD and AIKS suggest that NormTrunc with $k =$ 10 provides high quality samples. It is observed that the NormTrunc sample with $k =$ 10 provides the same values of the summary statistics as the gold standard. On the other hand, the ESS-recommended NormTrunc sample with $k =$ 3 does not match the gold standard. It provides a large median and a high right-tail probability compared to the gold standard. In summary, our approaches help tune algorithms. In particular, for asymptotically inexact algorithms, our methods can guide users to appropriately choose their tuning parameters and help provide reliable inference.

\section{Discussion}
\label{sec:discussion}

In this article we proposed new methods that provide guidance for tuning algorithms and give some measure of sample quality for a wide range of Monte Carlo algorithms, including a particularly challenging class of algorithms: asymptotically inexact algorithms for distributions with intractable normalizing functions. We describe three methods. CD applies broadly to most any likelihood-based context where misspecification is of concern while ACD and AIKS are specifically targeted at likelihoods with intractable normalizing functions. Our study mainly focuses on intractable normalizing function problems and shows that our methods can assess the quality of samples and provide good guidance for tuning algorithms. We have studied applications of ACD and AIKS to several asymptotically exact and inexact algorithms in the context of challenging simulated and real data examples. This shows that our methods provide useful results not only for asymptotically exact algorithms, for which some other methods may be useful, but also for asymptotically inexact algorithms, for which we are not aware of other methods. 

There are of course simple and reasonable heuristics one can use for diagnostic purposes, such as increasing $m$ until DMH approximations stabilize. However, our diagnostics go beyond simple approaches, allowing one to compare sample quality across different algorithms, including comparing asymptotically exact algorithms with asymptotically inexact algorithms.

We note that ACD and AIKS could be slightly different in their conclusions, as our examples illustrate, but the difference is quite small and conclusions are nearly the same overall. 
The difference between ACD and AIKS can stem from the fact that they consider different sets of test functions to quantify the difference between a sample mean and a target expectation. ACD considers two functions whose target expectations are identical and measures the difference between sample means of the two functions. AIKS considers a set of functions whose target expectations are zero and finds the maximum error between the sample mean and the target expectation. Unless the score function of the target density is the origin, their sets of test functions are mutually exclusive, which is the case of all the examples in this article. 


ACD and AIKS are obtained by replacing some functions that have to be evaluated for CD and IMQ KSD with their Monte Carlo approximations. This allows for the diagnostics to be available for doubly intractable posterior distributions. The computing time of ACD and AIKS mainly accounts for the Monte Carlo approximation, which might be computationally expensive for high-dimensional datasets. However, the approximation step is embarrassingly parallel and the computational cost can be reduced drastically via the self-normalized importance sampling.
An important caveat of ACD is that it can be misled if the first two moments match the target distribution but higher order moments do not. AIKS cannot be misled in such fashion. 
Extending ACD to higher order moments may provide interesting avenues for future research.

\section*{Acknowledgements}
The authors are grateful to Jaewoo Park for providing useful code and to Galin Jones and Geoff K. Nicholls for illuminating discussions. JH and MH were partially supported by the National Institute of General Medical Sciences of the National Institutes of Health under Award Number R01GM123007.


\appendix
\section{Monte Carlo approximations to intractable terms}
\label{app:sec:mcapprox}

\subsection{Derivative of log normalizing function}
\label{subsec:scorec}
For a $p$-dimensional parameter vector $\bth$, consider a posterior density $\pi(\bth \mid \bx)$ whose likelihood function is $L(\bth \mid \bx) = h(\bx \mid \bth)/c(\bth)$ and prior density is $p(\bth)$. The $k$th entry of $\nabla_{\bth} \log c(\bth \mid \bx)$, the score function of the normalizing function, can be written as 
\begin{align}
    \frac{\partial \log c(\bth)}{\partial \theta_k} &= \frac{1}{c(\bth)} \frac{\partial c(\bth)}{\partial \theta_k} \nonumber\\
    &= \frac{1}{c(\bth)} \frac{\partial }{\partial \theta_k} \int_{\boldsymbol{\mathcal{X}}} h(\bx | \bth) d\bx \nonumber\\
    &= \frac{1}{c(\bth)} \int_{\boldsymbol{\mathcal{X}}} \frac{\partial h(\bx | \bth)}{\partial \theta_k} d\bx \label{eq:dct} \\
    &= \frac{1}{c(\bth)} \int_{\boldsymbol{\mathcal{X}}} h(\bx | \bth) \frac{\partial \log h(\bx | \bth)}{\partial \theta_k} d\bx \nonumber \\
    &= \int_{\boldsymbol{\mathcal{X}}}  \frac{\partial \log h(\bx | \bth)}{\partial \theta_k} f(\bx | \bth) d\bx \nonumber \\
    &= \textup{E}_{f} \left\{ \frac{\partial \log h(\bX | \bth)}{\partial \theta_k} \right\}. \label{eq:scorec}
\end{align}
Equation~\eqref{eq:dct} follows from the dominated convergence theorem. Under the assumptions that the score function exists and the score function and the normalizing function are bounded, we have exchanged the derivative with the integral.

\subsection{Second derivative of log normalizing function}
\label{subsec:hessianc}
The $(k,l)$th entry of $\frac{\partial}{\partial \bth}\nabla_{\bth} \log c(\bth \mid \bx)$, the Hessian matrix of the normalizing function, is given by 
\begin{align}
    \frac{\partial^2 \log c(\bth)}{\partial \theta_k \partial \theta_l} &= \frac{\partial}{\partial \theta_k} \left\{ \frac{1}{c(\bth)} \frac{\partial c(\bth)}{\partial \theta_l} \right\} \nonumber\\
    &= \frac{1}{c(\bth)} \frac{\partial^2 c(\bth)}{\partial \theta_k \partial \theta_l} - \left\{ \frac{1}{c(\bth)} \frac{\partial c(\bth)}{\partial \theta_k} \right\} \left\{ \frac{1}{c(\bth)} \frac{\partial c(\bth)}{\partial \theta_l} \right\} \nonumber\\
    &= \frac{1}{c(\bth)} \frac{\partial^2 c(\bth)}{\partial \theta_k \partial \theta_l} -  \textup{E}_{f} \left\{ \frac{\partial \log h(\bX | \bth)}{\partial \theta_k} \right\}  \textup{E}_{f} \left\{ \frac{\partial \log h(\bX | \bth)}{\partial \theta_l} \right\}. \label{eq:seconddervc}
\end{align}
Now consider the first term on the right hand side of equation~\eqref{eq:seconddervc}. It can be written as
\begin{align}
    \frac{1}{c(\bth)} \frac{\partial^2 c(\bth)}{\partial \theta_k \partial \theta_l}  &=\frac{1}{c(\bth)} \frac{\partial^2}{\partial \theta_k \partial \theta_l} \int_{\bs{\mathcal{X}}} h(\bx | \bth) d\bx \nonumber \\
    &= \frac{1}{c(\bth)} \int_{\bs{\mathcal{X}}} \frac{\partial}{\partial \theta_k} \left\{ \frac{\partial \log h(\bx | \bth)}{\partial \theta_l} h(\bx | \bth) \right\} d\bx \label{eq:dctHess}\\
    &=\int_{\bs{\mathcal{X}}} \frac{\partial^2 \log h(\bx | \bth)}{\partial \theta_k \partial \theta_l} f(\bx | \bth) d\bx + \int_{\bs{\mathcal{X}}} \frac{\partial \log h(\bx | \bth)}{\partial \theta_k}  \frac{\partial \log h(\bx | \bth)}{\partial \theta_l} f(\bx | \bth) d\bx \nonumber \\
    &=\textup{E}_{f} \left\{ \frac{\partial^2 \log h(\bX | \bth)}{\partial \theta_k \partial \theta_l} \right\} + \textup{E}_{f} \left\{ \frac{\partial \log h(\bX | \bth)}{\partial \theta_k} \frac{\partial \log h(\bX | \bth)}{\partial \theta_l} \right\}. \label{eq:seconddervc2}
\end{align}
In the equation~\eqref{eq:dctHess} we have exchanged the derivative with the integral, owing to the dominated convergence theorem. Combining \eqref{eq:seconddervc} and \eqref{eq:seconddervc2}, the entry of $\frac{\partial}{\partial \bth}\nabla_{\bth} \log c(\bth \mid \bx)$ is 
\begin{align}
    \frac{\partial^2 \log c(\bth)}{\partial \theta_k \partial \theta_l} &=\textup{E}_{f} \left\{ \frac{\partial^2 \log h(\bX | \bth)}{\partial \theta_k \partial \theta_l} \right\} + \textup{E}_{f} \left\{ \frac{\partial \log h(\bX | \bth)}{\partial \theta_k} \frac{\partial \log h(\bX | \bth)}{\partial \theta_l} \right\} \nonumber \\ & -  \textup{E}_{f} \left\{ \frac{\partial \log h(\bX | \bth)}{\partial \theta_k} \right\}  \textup{E}_{f} \left\{ \frac{\partial \log h(\bX | \bth)}{\partial \theta_l} \right\}. \label{eq:seconddervc3}
\end{align}

\subsection{Monte Carlo approximations}

We approximate the $k$th entry of the vector $\nabla_{\bth} \log c(\bth)$ using a sample generated from the model distribution:
\begin{align}
    \frac{\partial \log c(\bth)}{\partial \theta_k} &\approx \frac{1}{N} \sum_{j=1}^{N}  \frac{ \partial \log h(\by^{(j)}|\bth)}{\partial \theta_k}, \nonumber
\end{align}
where $\by^{(1)}, \dots, \by^{(N)}$ are auxiliary variables generated exactly from $f(\cdot \mid \bth)$ or generated by a Monte Carlo algorithm having $f(\cdot \mid \bth)$ as its target distribution. 
In a similar fashion, the approximation of the $(k,l)$th entry of $\left\{\nabla_{\bth}\log c(\bth)\right\}$ $\left\{\nabla_{\bth} \log c(\bth)\right\}^\top$ is calculated as
\begin{align*}
    \frac{\partial \log c(\bth)}{\partial \theta_k} \frac{\partial \log c(\bth)}{\partial \theta_l} &\approx \left\{ \frac{1}{N} \sum_{j=1}^{N}  \frac{ \partial \log h(\by^{(j)}|\bth)}{\partial \theta_k} \right\} \left\{ \frac{1}{N} \sum_{j=1}^{N}  \frac{ \partial \log h(\by^{(j)}|\bth)}{\partial \theta_l} \right\}, \; k, l = 1,\dots,p.
\end{align*}
Lastly we approximate $(k,l)$th entry of $\frac{\partial}{\partial \bth} \nabla_{\bth} \log c(\bth)$ by
\begin{align*}
    \frac{\partial^2 \log c(\bth)}{\partial \theta_k \theta_l} &\approx \frac{1}{N} \sum_{j=1}^{N} \frac{\partial^2 \log h(\by^{(j)} | \bth)}{\partial \theta_k \partial \theta_l} + \frac{1}{N} \sum_{j=1}^{N}  \frac{\partial \log h(\by^{(j)}|\bth)}{\partial \theta_k}   \frac{\partial \log h(\by^{(j)}|\bth)}{\partial \theta_l} \\
    & - \left\{ \frac{1}{N} \sum_{j=1}^{N} \frac{\partial \log h(\by^{(j)}|\bth)}{\partial \theta_k} \right\} \left\{ \frac{1}{N} \sum_{j=1}^{N} \frac{\partial \log h(\by^{(j)}|\bth)}{\partial \theta_l} \right\}.
\end{align*}

\section{Proof of Theorem~\ref{thm:ACD}}
\label{app:sec:thmACD}

Let $D(\bth) = J(\bth) + H(\bth)$ and $D_{k,l}(\bth)$ be the $(k,l)$ entry of $D(\bth)$. Let $\hat{D}_{k,l}(\bth)$ be the approximation, computed using auxiliary variables $\by^{(1)}$,\dots,$\by^{(N)}$ generated from the model distribution, to $D_{k,l}(\bth)$. Suppose we have an i.i.d. sample $\bth^{(1)}, \dots, \bth^{(n)}$ generated from $\pi(\cdot \mid \bx)$. For all $k,l,m,s \in \{1, \dots, p\}$, the difference between an entry of $ V = \textup{E}_{\pi} \{ \bs{d}(\bth) \bs{d}(\bth)^\top \}$ and its two-state approximation is
\begin{align}
    &\left \vert \textup{E}_{\pi} \{ D_{k,l}(\bth) D_{m,s}(\bth) \} - \frac{1}{n} \sum_{i=1}^n \hat{D}_{k,l}(\bth^{(i)}) \hat{D}_{m,s}(\bth^{(i)}) \right \vert \nonumber \\
    & \leq \left \vert \frac{1}{n} \sum_{i=1}^n D_{k,l}(\bth^{(i)}) D_{m,s}(\bth^{(i)}) - \textup{E}_{\pi} \{ D_{k,l}(\bth) D_{m,s}(\bth) \}  \right \vert \nonumber \\
    & \quad + \left \vert \frac{1}{n} \sum_{i=1}^n \left \{ \hat{D}_{k,l}(\bth^{(i)}) \hat{D}_{m,s}(\bth^{(i)}) - D_{k,l}(\bth^{(i)}) D_{m,s}(\bth^{(i)}) \right \} \right \vert \nonumber \\
    & \leq \delta(n) +  \frac{1}{n} \sum_{i=1}^n \left \vert \hat{D}_{k,l}(\bth^{(i)}) \hat{D}_{m,s}(\bth^{(i)}) - D_{k,l}(\bth^{(i)}) D_{m,s}(\bth^{(i)}) \right \vert, \label{ineq:dd}
\end{align}
where $\delta(n) = \Vert \frac{1}{n} \sum_{i=1}^n \bs{d}(\bth^{(i)}) \bs{d}(\bth^{(i)})^\top - \textup{E}_{\pi} \{ \bs{d}(\bth) \bs{d}(\bth)^\top \} \Vert_{\max} = \mathcal{O}(n^{-1/2})$ from the ergodic theorem. Now we consider the second term on the right hand side of inequality~\eqref{ineq:dd}. Let $u_k(\bth)$ be the $k$th entry of the score function $\bs{u}(\bth)$ of the posterior distribution and $\hat{u}_k(\bth)$ be its approximation computed using the auxiliary variables. Let $H_{k,l}(\bth)$ be the $(k,l)$th entry of the hessian matrix $H(\bth)$ of the posterior distribution and $\hat{H}_{k,l}(\bth)$ be its approximation computed using the auxiliary variables. We can bound the difference between $\hat{D}_{k,l}(\bth) \hat{D}_{m,s}(\bth)$ and $D_{k,l}(\bth) D_{m,s}(\bth)$ as follows. For all $k,l,m,s \in \{1, \dots, p\}$, we have
\begin{align}
    \left \vert \hat{D}_{k,l}(\bth) \hat{D}_{m,s}(\bth) - D_{k,l}(\bth) D_{m,s}(\bth) \right \vert & \leq  \left \vert \hat{u}_k(\bth) \hat{u}_l (\bth) \hat{u}_m(\bth) \hat{u}_s (\bth) - u_k(\bth) u_l (\bth) u_m(\bth) u_s (\bth) \right \vert \nonumber \\
    & \quad + \left \vert \hat{u}_k(\bth) \hat{u}_l (\bth) \hat{H}_{m,s}(\bth) - u_k(\bth) u_l (\bth) H_{m,s}(\bth) \right \vert \nonumber \\
    & \quad + \left \vert \hat{u}_m(\bth) \hat{u}_s (\bth) \hat{H}_{k,l}(\bth) - u_m(\bth) u_s (\bth) H_{k,l}(\bth) \right \vert \nonumber \\
    & \quad + \left \vert  \hat{H}_{k,l}(\bth)  \hat{H}_{m,s}(\bth) -  H_{k,l}(\bth)  H_{m,s}(\bth) \right \vert. \label{ineq:dhatdahtmdd}
\end{align}
 
Let 
\begin{align*}
    \epsilon_1(N) &= \max_k \left| \frac{1}{N} \sum_{j=1}^N\frac{\partial \log h(\by^{(j)} | \bth)}{\partial \theta_k} - \textup{E}_f \left\{ \frac{\partial \log h(\bX | \bth)}{\partial \theta_k} \right\} \right|,\\
   \epsilon_2(N)  &= \max_{k,l} \left| \frac{1}{N} \sum_{j=1}^{N} \frac{\partial^2 \log h(\by^{(j)} | \bth)}{\partial \theta_k \partial \theta_l}  - \textup{E}_f \left\{ \frac{\partial^2 \log h(\bX | \bth)}{\partial \theta_k\partial \theta_l}  \right\}  \right|,\\
   \epsilon_3(N) &= \max_l \left| \frac{1}{N} \sum_{j=1}^{N}\frac{\partial^3 \log h(\by^{(j)} | \bth)}{\partial \theta_k \partial \theta_l \partial \theta_m} - \textup{E}_f \left\{ \frac{\partial^3 \log h(\bX | \bth)}{\partial \theta_k \partial \theta_l \partial \theta_m} \right\} \right|,\\
   \epsilon_4(N) &= \max_{k,l} \left| \frac{1}{N} \sum_{j=1}^{N} \left\{ \frac{\partial \log h(\by^{(j)}|\bth)}{\partial \theta_k} \right\} \left\{ \frac{\partial \log h(\by^{(j)}|\bth)}{\partial \theta_l} \right\} - \textup{E}_f \left\{  \frac{\partial \log h(\bX |\bth)}{\partial \theta_k}  \frac{\partial \log h(\bX |\bth)}{\partial \theta_l}  \right\} \right|\\
   \epsilon_5(N) &= \max_{k,l,m} \left| \frac{1}{N} \sum_{j=1}^{N} \left\{ \frac{\partial \log h(\by^{(j)}|\bth)}{\partial \theta_k} \right\} \left\{ \frac{\partial^2 \log h(\by^{(j)}|\bth)}{\partial \theta_l \partial \theta_m} \right\} - \textup{E}_f \left\{  \frac{\partial \log h(\bX |\bth)}{\partial \theta_k}  \frac{\partial^2 \log h(\bX |\bth)}{\partial \theta_l \partial \theta_m} \right\} \right|,\\
   \epsilon_6(N) &= \max_{k,l,m} \Bigg| \frac{1}{N} \sum_{j=1}^{N} \left\{ \frac{\partial \log h(\by^{(j)}|\bth)}{\partial \theta_k} \right\} \left\{ \frac{\partial \log h(\by^{(j)}|\bth)}{\partial \theta_l} \right\} \left\{ \frac{\partial \log h(\by^{(j)}|\bth)}{\partial \theta_m} \right\} \\
   & \qquad \qquad - \textup{E}_f \left\{  \frac{\partial \log h(\bX |\bth)}{\partial \theta_k}  \frac{\partial \log h(\bX |\bth)}{\partial \theta_l} \frac{\partial \log h(\bX |\bth)}{\partial \theta_m}  \right\} \Bigg|, 
\end{align*}
and $\epsilon(N) = \max \left\{ \epsilon_1(N), \epsilon_2(N), \epsilon_3(N), \epsilon_4(N), \epsilon_5(N), \epsilon_6(N) \right\}$ = $\mathcal{O}(N^{-1/2})$ from the ergodic theorem. Now consider the first term on the right hand side of inequality~\eqref{ineq:dhatdahtmdd}. For all $k,l,m,s \in \{1, \dots, p\}$, we have
\begin{align}
    \left \vert \hat{u}_k(\bth) \hat{u}_l (\bth) \hat{u}_m(\bth) \hat{u}_s (\bth) - u_k(\bth) u_l (\bth) u_m(\bth) u_s (\bth) \right \vert \leq B_1 \epsilon(N), \nonumber
\end{align}
where $B_1$ is some constant. The above inequality follows from Assumptions 1 and 2, and the following elementary inequalities:
\begin{align}
    \vert ab - \hat{a}\hat{b} \vert &= \vert ab - \hat{a}b + \hat{a}b - \hat{a}\hat{b} \vert =  \vert (a - \hat{a})b + (b - \hat{b})\hat{a} \vert \nonumber \\ 
    & \leq \vert a - \hat{a}\vert \vert b \vert + \vert b - \hat{b}\vert \vert\hat{a}\vert, \label{ineq:eleIneq} \\
    \vert abc - \hat{a}\hat{b}\hat{c} \vert &= \vert abc - \hat{a}bc + \hat{a}bc - \hat{a}\hat{b}\hat{c} \vert = \vert (a-\hat{a})bc + (bc - \hat{b}\hat{c})\hat{a} \vert \nonumber \\
    & \leq \vert a - \hat{a} \vert \vert b \vert \vert c \vert + \vert bc - \hat{b}\hat{c} \vert \vert \hat{a} \vert \label{ineq:eleIneq2}, \\
    \vert abcd - \hat{a}\hat{b}\hat{c}\hat{d} \vert &= \vert abcd - \hat{a}bcd + \hat{a}bcd - \hat{a}\hat{b}\hat{c}\hat{d}  \vert = \vert (a - \hat{a})bcd + (bcd - \hat{b}\hat{c}\hat{d})\hat{a} \vert \nonumber \\
    & \leq \vert a - \hat{a} \vert \vert b \vert \vert c \vert \vert d \vert + \vert bcd - \hat{b}\hat{c}\hat{d} \vert \vert \hat{a} \vert. \label{ineq:eleIneq3}
\end{align}

Now consider the remaining terms on the right hand side of inequality~\eqref{ineq:dhatdahtmdd}. In a similar fashion, for all $k,l,m,s \in \{1, \dots, p\}$, we have
\begin{align*}
    \left \vert \hat{u}_k(\bth) \hat{u}_l (\bth) \hat{H}_{m,s}(\bth) - u_k(\bth) u_l (\bth) H_{m,s}(\bth) \right \vert & \leq B_2 \epsilon(N),\\
    \left \vert \hat{u}_m(\bth) \hat{u}_s (\bth) \hat{H}_{k,l}(\bth) - u_m(\bth) u_s (\bth) H_{k,l}(\bth) \right \vert &\leq B_3 \epsilon(N),\\
    \vert \hat{H}_{k,l}(\bth) \hat{H}_{m,s}(\bth) - H_{k,l}(\bth) H_{m,s}(\bth) \vert &\leq B_4 \epsilon(N), 
\end{align*}
where $B_2, B_3$, and $B_4$ are some constants. These inequalities follow from Assumptions 1 and 2 in the manuscript and inequalities \eqref{ineq:eleIneq} to \eqref{ineq:eleIneq3}. Therefore, the approximation error of the two-stage approximation to the asymptotic covariance matrix is 
\begin{align}
    \Vert \hat{V}_{n,N} - V \Vert_{\max} &\leq \mathcal{O}\left( n^{-1/2} \right) + \mathcal{O}\left( N^{-1/2} \right), \nonumber
\end{align}
almost surely. We use $\Vert \cdot \Vert_{\max}$ to represent the max norm of vectors or matrices.

\section{Proof of Theorem~\ref{thm:ACD_bm}}
\label{app:sec:thmACDbm}

Suppose we have a sample $\bth^{(1)}, \dots, \bth^{(n)}$ generated from a polynomially ergodic Markov chain with $\pi(\bth \mid \bx)$ as its stationary distribution. Let $D(\bth) = J(\bth) + H(\bth)$ and $\hat{D}(\bth)$ be the approximation, computed using auxiliary variables $\by^{(1)}$,\dots,$\by^{(N)}$ generated from the model distribution, to $D(\bth)$. Let $\bar{D} = \frac{1}{n} \sum_{i=1}^{n}D(\bth^{(i)})$, $\hat{\bar{D}} = \frac{1}{n} \sum_{i=1}^{n}\hat{D}(\bth^{(i)})$, $\Bar{D}^j = \frac{1}{b_n} \sum_{i=(j-1)b_n+1}^{jb_n} D(\bth^{(i)})$, and $\hat{\Bar{D}}^j = \frac{1}{b_n} \sum_{i=(j-1)b_n+1}^{jb_n} \hat{D}(\bth^{(i)})$ for $j = 1, \dots, a_n$, where $a_n$ is the number of batches and $b_n$ is the batch size. For all $k,l,m,s \in \{1, \dots, p\}$, the difference between an entry $\sigma_r
$ of $ \Sigma$ and its two-state batch means approximation is
\begin{align}
    &\left \vert \sigma_r - \frac{b_n}{a_n-1} \sum_{j=1}^{a_n} (\hat{\Bar{D}}^j_{k,l} - \hat{\Bar{D}}_{k,l})(\hat{\Bar{D}}^j_{m,s} - \hat{\Bar{D}}_{m,s}) \right \vert \nonumber \\
    & \leq \left \vert \frac{b_n}{a_n-1} \sum_{j=1}^{a_n} (\Bar{D}^j_{k,l} - \Bar{D}_{k,l})(\Bar{D}^j_{m,s} - \Bar{D}_{m,s}) - \sigma_r \right \vert \nonumber \\
    & \quad + \left \vert \frac{b_n}{a_n-1} \sum_{j=1}^{a_n} \left \{ (\hat{\Bar{D}}^j_{k,l} - \hat{\Bar{D}}_{k,l})(\hat{\Bar{D}}^j_{m,s} - \hat{\Bar{D}}_{m,s}) - (\Bar{D}^j_{k,l} - \Bar{D}_{k,l})(\Bar{D}^j_{m,s} - \Bar{D}_{m,s}) \right \} \right \vert \nonumber \\
    & \leq \gamma(n) +  \frac{b_n}{a_n-1} \sum_{j=1}^{a_n} \left \vert (\hat{\Bar{D}}^j_{k,l} - \hat{\Bar{D}}_{k,l})(\hat{\Bar{D}}^j_{m,s} - \hat{\Bar{D}}_{m,s}) - (\Bar{D}^j_{k,l} - \Bar{D}_{k,l})(\Bar{D}^j_{m,s} - \Bar{D}_{m,s}) \right \vert, \label{ineq:dd_batch}
\end{align}
where $\gamma(n) = \Vert \hat{\Sigma}_{n} - \Sigma \Vert_{\max} = \mathcal{O}\left( (\log n / b_n)^{1/2} n^{1/2-\lambda}\right) \to 0$ as $n \to \infty$ for some $\lambda > 0$ under some conditions \citep[Theorem 2 in][]{Vats2019}. Now we consider the second term on the right hand side of inequality~\eqref{ineq:dd_batch}. For all $k,l,m,s \in \{1, \dots, p\}$, we have

\begin{align}
    &b_n \left \vert(\hat{\Bar{D}}^j_{k,l} - \hat{\Bar{D}}_{k,l}) (\hat{\Bar{D}}^j_{m,s} - \hat{\Bar{D}}_{m,s}) - (\Bar{D}^j_{k,l} - \Bar{D}_{k,l}) (\Bar{D}^j_{m,s} - \Bar{D}_{m,s}) \right \vert \nonumber \\
    & = \left \vert b_n^{1/2}(\hat{\Bar{D}}^j_{k,l} - \hat{\Bar{D}}_{k,l}) b_n^{1/2}(\hat{\Bar{D}}^j_{m,s} - \hat{\Bar{D}}_{m,s}) - b_n^{1/2}(\Bar{D}^j_{k,l} - \Bar{D}_{k,l}) b_n^{1/2}(\Bar{D}^j_{m,s} - \Bar{D}_{m,s}) \right \vert \nonumber \\
    & \leq \left \vert b_n^{1/2}(\Bar{D}^j_{k,l} - \Bar{D}_{k,l}) - b_n^{1/2}(\hat{\Bar{D}}^j_{k,l} - \hat{\Bar{D}}_{k,l}) \right \vert \left\vert b_n^{1/2}(\Bar{D}^j_{m,s} - \Bar{D}_{m,s})  \right\vert \nonumber \\
    & \quad + \left\vert b_n^{1/2}(\Bar{D}^j_{m,s} - \Bar{D}_{m,s}) - b_n^{1/2}(\hat{\Bar{D}}^j_{m,s} - \hat{\Bar{D}}_{m,s}) \right\vert \left\vert b_n^{1/2}(\hat{\Bar{D}}^j_{k,l} - \hat{\Bar{D}}_{k,l}) \right\vert \label{ineq:ab_ahatbhat_batch} \\
    & \leq \left\vert b_n^{1/2}(\Bar{D}^j_{k,l} - \hat{\Bar{D}}^j_{k,l}) - b_n^{1/2} (\Bar{D}_{k,l} - \hat{\Bar{D}}_{k,l} ) \right\vert \nonumber \\
    & \quad \times \left\vert b_n^{1/2} \left\{ \Bar{D}^j_{m,s} - \textup{E}_{\pi}[D_{m,s}(\bth)] \right\} - b_n^{1/2} \left\{ \Bar{D}_{m,s} - \textup{E}_{\pi}[D_{m,s}(\bth)] \right\}
    \right\vert \nonumber \\
    & \quad + \left\vert b_n^{1/2}(\Bar{D}^j_{m,s} - \hat{\Bar{D}}^j_{m,s}) - b_n^{1/2}(\Bar{D}_{m,s} - \hat{\Bar{D}}_{m,s}) \right\vert \nonumber \\ 
    & \quad \times \left\vert b_n^{1/2} (\hat{\Bar{D}}^j_{k,l} - \Bar{D}^j_{k,l}) - b_n^{1/2}(\hat{\Bar{D}}_{k,l} - \Bar{D}_{k,l}) +  b_n^{1/2} \left\{ \Bar{D}^j_{k,l} - \textup{E}_{\pi}[D_{k,l}(\bth)] \right\} - b_n^{1/2} \left\{ \Bar{D}_{k,l} - \textup{E}_{\pi}[D_{k,l}(\bth)] \right\} \right\vert \nonumber \\
    & \leq \left\{ b_n^{1/2} \vert \Bar{D}^j_{k,l} - \hat{\Bar{D}}^j_{k,l} \vert + b_n^{1/2} \vert \Bar{D}_{k,l} - \hat{\Bar{D}}_{k,l} \vert \right\} \nonumber \\
    & \quad \times \left\{ b_n^{1/2} \left \vert \Bar{D}^j_{m,s} - \textup{E}_{\pi}[D_{m,s}(\bth)] \right \vert + b_n^{1/2} \left \vert \Bar{D}_{m,s} - \textup{E}_{\pi}[D_{m,s}(\bth)] \right \vert \right\} \nonumber \\
    & \quad + \left\{ b_n^{1/2} \vert \Bar{D}^j_{m,s} - \hat{\Bar{D}}^j_{m,s} \vert + b_n^{1/2} \vert \Bar{D}_{m,s} - \hat{\Bar{D}}_{m,s} \vert \right\} \nonumber \\
    & \quad \times \left\{ b_n^{1/2} \vert \Bar{D}^j_{k,l} - \hat{\Bar{D}}^j_{k,l} \vert + b_n^{1/2} \vert \Bar{D}_{k,l} - \hat{\Bar{D}}_{k,l} \vert + b_n^{1/2} \left\vert \Bar{D}^j_{k,l}- \textup{E}_{\pi}[D_{k,l}(\bth)] \right\vert + b_n^{1/2} \left\vert \Bar{D}_{k,l} - \textup{E}_{\pi}[D_{k,l}(\bth)] \right \vert \right\} \nonumber\\
    &\leq \left\{ b_n^{1/2} \mathcal{O}\left(N^{-1/2}\right) + b_n^{1/2} \mathcal{O}\left(N^{-1/2}\right) \right\} \left\{ b_n^{1/2} \mathcal{O}\left(b_n^{-1/2}\right) + b_n^{1/2} \mathcal{O}\left(n^{-1/2}\right) \right\} \nonumber \\
    & \quad + \left\{ b_n^{1/2} \mathcal{O}\left(N^{-1/2}\right) + b_n^{1/2} \mathcal{O}\left(N^{-1/2}\right) \right\} \nonumber \\
    & \quad \times \left\{ b_n^{1/2} \mathcal{O}\left(N^{-1/2}\right) + b_n^{1/2} \mathcal{O}\left(N^{-1/2}\right) + b_n^{1/2} \mathcal{O}\left(b_n^{-1/2}\right) + b_n^{1/2} \mathcal{O}\left(n^{-1/2}\right) \right\} \nonumber \\
    & \leq \mathcal{O}\left( (b_n/N)^{1/2}\right) + \mathcal{O}\left((b_n/n)^{1/2} (b_n/N)^{1/2} \right) + \mathcal{O}\left( b_n/N \right) = \mathcal{O}\left( (b_n/N)^{1/2} \right), \nonumber
\end{align}
where inequality~\eqref{ineq:ab_ahatbhat_batch} follows from inequality~\eqref{ineq:eleIneq}. Therefore, the approximation error of the two-stage batch means approximation to the asymptotic covariance matrix is
\begin{align}
    \Vert \hat{\Sigma}_{n,N} - \Sigma \Vert_{\max} &\leq \mathcal{O}\left( (\log n/b_n)^{1/2} n^{1/2 - \lambda} \right) + \mathcal{O}\left((b_n/N)^{1/2}\right), \nonumber
\end{align}
almost surely for some $\lambda > 0$.

\section{Proof of Theorem~\ref{thm:AIKS}}
\label{app::sec:thmAIKS}
Consider a target distribution $P$ and a weighted sample $Q_n = \sum_{i=1}^n q_n(\bth^{(i)}) \delta_{\bth^{(i)}}$ targeting $P$, where $\bth^{(1)}, \dots, \bth^{(n)}$ are sample points and $q_n$ is a probability mass function. By Minkowski's inequality, the difference between IMQ KSD and AIKS of $Q_n$ is
\begin{align}
    \left| \hat{\mathcal{S}}(Q_n, \mathcal{T}_P, \mathcal{G}_{k, \Vert \cdot \Vert_p} )- \mathcal{S}(Q_n, \mathcal{T}_P, \mathcal{G}_{k, \Vert \cdot \Vert_p}) \right| &\leq \Vert \hat{\bw} - \bw \Vert_p. \nonumber 
\end{align}
We now consider the term on the right hand side of this inequality.
\begin{align}
    \Vert \hat{\bw} - \bw \Vert_p^p &= \sum_{j=1}^p |\hat{w}_j - w_j|^p \nonumber \\
    &= \sum_{j=1}^p \frac{|\hat{w}_j - w_j|^{p-1}}{\hat{w}_j + w_j} |\hat{w}_j^2 - w_j^2| \nonumber \\
    &= \sum_{j=1}^p \frac{|\hat{w}_j - w_j|^{p-1}}{\hat{w}_j + w_j} \sum_{k,l = 1}^n q_n(\bth^{(k)})q_n(\bth^{(l)}) \left| \hat{k}_0^j(\bth^{(k)}, \bth^{(l)}) - k_0^j(\bth^{(k)}, \bth^{(l)}) \right|. \nonumber
\end{align}
Let $\epsilon(N) = \max_{k} \Vert \hat{\bu}(\bth^{(k)}) -  \bu(\bth^{(k)}) \Vert_{\max} = \mathcal{O}(1/\sqrt{N})$ from ergodic theorem. The approximate error for the Stein kernel can be derived as follows.
\begin{align}
    & \left| \hat{k}_0^j(\bth^{(k)}, \bth^{(l)}) - k_0^j(\bth^{(k)}, \bth^{(l)}) \right| \nonumber \\
    &\leq k(\bth^{(k)}, \bth^{(l)}) \left| \hat{u}_j(\bth^{(k)}) \hat{u}_j(\bth^{(l)}) - u_j(\bth^{(k)}) u_j(\bth^{(l)}) \right| \nonumber \\
    & + \left| \nabla_{\theta_j^{(l)}}  k(\bth^{(k)}, \bth^{(l)}) \right| \left| \hat{u}_j(\bth^{(k)}) - u_j(\bth^{(k)}) \right| \nonumber \\
    & + \left| \nabla_{\theta_j^{(k)}}  k(\bth^{(l)}, \bth^{(k)}) \right| \left| \hat{u}_j(\bth^{(l)}) - u_j(\bth^{(l)}) \right| \nonumber \\
    &\leq \left[ k(\bth^{(k)}, \bth^{(l)}) \left\{ |\hat{u}_j(\bth^{(k)})| + |u_j(\bth^{(l)})| \right\} + 2 | \nabla_{\theta_j^{(l)}}  k(\bth^{(k)}, \bth^{(l)}) | \right] \epsilon(N) \label{ineq:th2ab} \\
    &\leq c_1 \epsilon(N) \nonumber
\end{align}
for bounded constant $c_1$. The inequality in~\eqref{ineq:th2ab} follows from \eqref{ineq:eleIneq} and the fact that $ \nabla_{\theta_j^{(l)}}  k(\bth^{(k)}, \bth^{(l)})$ and $ \nabla_{\theta_j^{(k)}}  k(\bth^{(l)}, \bth^{(k)})$ are symmetric around zero. Now, 
\begin{align}
    \Vert \hat{\bw} - \bw \Vert_p^p &\leq c_1 \epsilon(N) \sum_{j=1}^p \frac{|\hat{w}_j - w_j|^{p-1}}{\hat{w}_j + w_j} \sum_{k,l = 1}^n q_n(\bth^{(k)})q_n(\bth^{(l)})  \nonumber \\
    &\leq n^2 c_1 c_2 \epsilon(N) \nonumber
\end{align}
for bounded constant $c_2$. Therefore, the approximate error for AIKS is
\begin{align}
    \left| \hat{\mathcal{S}}(Q_n, \mathcal{T}_P, \mathcal{G}_{k, \Vert \cdot \Vert_p} )- \mathcal{S}(Q_n, \mathcal{T}_P, \mathcal{G}_{k, \Vert \cdot \Vert_p}) \right| &\leq \mathcal{O}\left( N^{-1/(2p)} \right) \nonumber 
\end{align}
almost surely.

\section{Proof of Proposition~\ref{prop:AIKS}}
\label{app::sec:propAIKS}

From theorem~\ref{thm:AIKS}, we have
    \begin{align*}
        \hat{\mathcal{S}}(Q_n, \mathcal{T}_P, \mathcal{G}_{k, \Vert \cdot \Vert_p}) &\leq \left\vert \hat{\mathcal{S}}(Q_n, \mathcal{T}_P, \mathcal{G}_{k, \Vert \cdot \Vert_p}) - \mathcal{S}(Q_n, \mathcal{T}_P, \mathcal{G}_{k, \Vert \cdot \Vert_p}) + \mathcal{S}(Q_n, \mathcal{T}_P, \mathcal{G}_{k, \Vert \cdot \Vert_p}) \right\vert\\
        &\leq \mathcal{O} \left\{ N^{-1/(2p)} \right\} + \mathcal{S}(Q_n, \mathcal{T}_P, \mathcal{G}_{k, \Vert \cdot \Vert_p}).
    \end{align*}
    The AIKS $\hat{\mathcal{S}}(Q_n, \mathcal{T}_P, \mathcal{G}_{k, \Vert \cdot \Vert_p})$ goes to zero when $Q_n \to P$ and $N \to \infty$. This implies that AIKS can detect convergence. Detecting non-convergence means that if AIKS goes to zero then $Q_n \to P$, equivalently, if $Q_n \not\to P$ then AIKS does not go to zero. From theorem~\ref{thm:AIKS} we have
    \begin{align*}
       \hat{\mathcal{S}}(Q_n, \mathcal{T}_P, \mathcal{G}_{k, \Vert \cdot \Vert_p}) & \geq \mathcal{S}(Q_n, \mathcal{T}_P, \mathcal{G}_{k, \Vert \cdot \Vert_p}) - \mathcal{O} \left\{ N^{-1/(2p)} \right\}
    \end{align*}
    The AIKS $\hat{\mathcal{S}}(Q_n, \mathcal{T}_P, \mathcal{G}_{k, \Vert \cdot \Vert_p})$ does not go to zero if $Q_n \not\to P$ and $N \to \infty$. This indicates that AIKS can detect non-convergence.

\section{Proof of Proposition~\ref{prop:asyDistAIKS}}
\label{app:sec:prop:asyDistAIKS}

We check the following assumptions from Theorem 2.1 of \citet{Leucht2013}. 
\begin{itemize}
    \item[A1] The process $\{\bth^{(i)}\}_{i \in \mathbb{Z}}$ is $\beta$-mixing with $\sum_{r=1}^\infty \sqrt{\beta(r)} < \infty$.
    \item[A2] Let $k_0 = \lim_{N \to \infty} \hat{k}_0$.
    \begin{itemize}
        \item[(i)] $k_0: \mathbb{R}^p \times \mathbb{R}^p \to \mathbb{R}$ is symmetric and degenerate.
        \item[(ii)] $k_0$ is positive semidefinite.
        \item[(iii)] $\text{E}_P\{ k_0 (\bth, \bth) \} < \infty$.
        \item[(iv)] $k_0$ is Lipschitz continuous.
    \end{itemize}
\end{itemize}
The conditions A1, A2 (iii), and A2 (iv) are assumed in Proposition~\ref{prop:asyDistAIKS}. 

The $k$ is symmetric, so is $k_0$. Let $Q$ be the stationary distribution of the process $\{\bth^{(i)}\}_{i \in \mathbb{Z}}$. We will show that $k_0$ is degenerate, i.e., $\text{E}_Q\{ k_0(\bth, \bth) \} = 0$ if and only if $Q = P$. If $Q = P$, then $\text{E}_Q\{ k_0 (\bth, \bth) \} = 0$ by Proposition 1 of \citet{Gorham2017}. Suppose that $Q \neq P$ but $\text{E}_Q\{ k_0 (\bth, \bth) \} = 0$. Since $\text{E}_Q\{ k_0 (\bth, \bth) \} = 0$, we have $\text{E}_Q \left\{ u_j(\bth) k(\bth, \cdot) + \frac{\partial k(\bth, \cdot)}{\partial \theta_j} \right\} = 0$ for $j = 1,\dots,p$ by the proof of Proposition 2 in \citet{Gorham2017}. Let $\pi$ and $q$ be the probability density functions of $P$ and $Q$, respectively. The $u_j(\bth) = \frac{\partial \log \pi(\bth) }{\partial \theta_j} = \frac{\partial \log q(\bth) }{\partial \theta_j} + \frac{\partial [ \log \pi(\bth) - \log q(\bth) ] }{\partial \theta_j}$. We have
\begin{align}
    \textbf{0} &= \textup{E}_Q \left\{ \nabla_{\bth} \log q(\bth) k(\bth, \cdot)  + \nabla_{\bth} [ \log \pi(\bth) - \log q(\bth) ] k(\bth, \cdot) + \nabla_{\bth} k(\bth, \cdot)  \right\} \nonumber\\
    &= \textup{E}_Q \left\{ \nabla_{\bth} \log q(\bth) k(\bth, \cdot) + \nabla_{\bth} k(\bth, \cdot)  \right\} + \textup{E}_Q \left\{ \nabla_{\bth} [ \log \pi(\bth) - \log q(\bth) ] k(\bth, \cdot) \right\} \nonumber \\
    &= \textup{E}_Q \left\{ \nabla_{\bth} [ \log \pi(\bth) - \log q(\bth) ] k(\bth, \cdot) \right\} \label{app:eq:embedding}
\end{align}
We note that the expectation of \eqref{app:eq:embedding} is the kernel embedding of a function $g(\bth) = \nabla_{\bth} \log \left( \frac{\pi(\bth)}{q(\bth)} \right)$ with respect to $Q$. By Theorem 4.4 c of \citet{carmeli2010vector}, the embedding is zero if and only if $g = 0$. Therefore, $\nabla_{\bth} \log \left( \frac{\pi(\bth)}{q(\bth)} \right) = \textbf{0}$ and thus $\log \left( \frac{\pi(\bth)}{q(\bth)} \right) = \textbf{c}$ where $\textbf{c}$ is a constant vector. This implies that $\pi(\bth) = e^{\textbf{c}} q(\bth)$. Since $\pi$ and $q$ integrate to one, $\textbf{c} = \textbf{0}$, which contradicts with $Q \neq P$.

The $k_0(\bth, \bth^\prime) = \langle \mathcal{T}_P k(\bth,\cdot), \mathcal{T}_P k(\bth^\prime,\cdot) \rangle_{\mathcal{K}_k}$ is an inner product and hence positive definite where $\mathcal{K}_k$ is a Hilbert space of functions such that, for all $\bth \in \mathbb{R}^p$, $k(\bth, \cdot) \in \mathcal{K}_k$ and $f(\bth) = \langle f, k(\bth, \cdot) \rangle_{\mathcal{K}_k}$ whenever $f \in \mathcal{K}_k$.


\section{Proof of Proposition~\ref{prop:bootAIKS}}
\label{app:sec:prop:bootAIKS}

We check assumptions from Theorem 3.1 of \citet{Leucht2013}. 
The condition B1 is assumed in Proposition~\ref{prop:bootAIKS}. The condition B2 is satisfied by Remark 2 (i) of \citet{Leucht2013}. The condition A2 follows from the proof of Proposition~\ref{prop:asyDistAIKS}.

\bibliographystyle{apalike}
\bibliography{refs}

\begin{thebibliography}{}

\bibitem[Alquier et~al., 2016]{alquier2016}
Alquier, P., Friel, N., Everitt, R., and Boland, A. (2016).
\newblock {Noisy Monte Carlo: Convergence of Markov chains with approximate
  transition kernels}.
\newblock {\em Statistics and Computing}, 26(1-2):29--47.

\bibitem[Andrieu and Roberts, 2009]{Andrieu2009}
Andrieu, C. and Roberts, G.~O. (2009).
\newblock {The pseudo-marginal approach for efficient Monte Carlo
  computations}.
\newblock {\em Annals of Statistics}, 37:697--725.

\bibitem[Atchad{\'{e}} et~al., 2013]{atchade2013}
Atchad{\'{e}}, Y.~F., Lartillot, N., and Robert, C. (2013).
\newblock {Bayesian computation for statistical models with intractable
  normalizing constants}.
\newblock {\em Brazilian Journal of Probability and Statistics},
  27(4):416--436.

\bibitem[Bartlett, 1953a]{Bartlett1953a}
Bartlett, M.~S. (1953a).
\newblock {Approximate confidence intervals}.
\newblock {\em Biometrika}, 40(1/2):12--19.

\bibitem[Bartlett, 1953b]{Bartlett1953b}
Bartlett, M.~S. (1953b).
\newblock {Approximate confidence intervals. II. More than one unknown
  parameter}.
\newblock {\em Biometrika}, 40(3/4):306--317.

\bibitem[Beaumont, 2003]{Beaumont2003}
Beaumont, M.~A. (2003).
\newblock {Estimation of population growth or decline in genetically monitored
  populations}.
\newblock {\em Genetics}, 164:1139--1160.

\bibitem[Besag, 1974]{Besag1974}
Besag, J. (1974).
\newblock {Spatial interaction and the statistical analysis of lattice
  systems}.
\newblock {\em Journal of the Royal Statistical Society: Series B},
  36(2):192--225.

\bibitem[Cameron and Johansson, 1997]{Cameron1997}
Cameron, A.~C. and Johansson, P. (1997).
\newblock {Count data regression using series expansions: With applications}.
\newblock {\em Journal of Applied Econometrics}, 12:203--223.

\bibitem[Carmeli et~al., 2010]{carmeli2010vector}
Carmeli, C., De~Vito, E., Toigo, A., and Umanit{\'a}, V. (2010).
\newblock {Vector valued reproducing kernel Hilbert spaces and universality}.
\newblock {\em Analysis and Applications}, 8(01):19--61.

\bibitem[Chanialidis et~al., 2018]{Chanialidis2018}
Chanialidis, C., Evers, L., Neocleous, T., and Nobile, A. (2018).
\newblock {Efficient Bayesian inference for COM-Poisson regression models}.
\newblock {\em Statistics and Computing}, 28:595--608.

\bibitem[Chazottes et~al., 2007]{Chazottes2007}
Chazottes, J.~R., Collet, P., Külske, C., and Redig, F. (2007).
\newblock {Concentration inequalities for random fields via coupling}.
\newblock {\em Probability Theory and Related Fields}, 137:201--225.

\bibitem[Chen and Seila, 1987]{Chen1987}
Chen, D. F.~R. and Seila, A.~F. (1987).
\newblock {Multivariate inference in stationary simulation using batch means}.
\newblock In {\em Proceedings of the 19th conference on Winter Simulation},
  pages 302--304.

\bibitem[Cheng et~al., 2017]{Cheng2017}
Cheng, J., Li, T., Levina, E., and Zhu, J. (2017).
\newblock {High-dimensional mixed graphical models}.
\newblock {\em Journal of Computational and Graphical Statistics}, 26:367--378.

\bibitem[Chwialkowski et~al., 2016]{Chwialkowski2016}
Chwialkowski, K., Strathmann, H., and Gretton, A. (2016).
\newblock {A kernel test of goodness of fit}.
\newblock In {\em Proceedings of the 33rd International Conference on Machine
  Learning}, volume~48, pages 2606--2615.

\bibitem[Conway and Maxwell, 1962]{Conway1962}
Conway, R.~W. and Maxwell, W.~L. (1962).
\newblock {Network dispatching by the shortest-operation discipline}.
\newblock {\em Operations Research}, 10:51--73.

\bibitem[Cowles and Carlin, 1996]{Cowles1996}
Cowles, M.~K. and Carlin, B.~P. (1996).
\newblock {Markov chain Monte Carlo convergence diagnostics: A comparative
  review}.
\newblock {\em Journal of the American Statistical Association},
  91(434):883--904.

\bibitem[Damerdji, 1994]{Damerdji1994}
Damerdji, H. (1994).
\newblock {Strong consistency of the variance estimator in steady-state
  simulation output analysis}.
\newblock {\em Mathematics of Operations Research}, 19:494--512.

\bibitem[Eddelbuettel and Francois, 2011]{Eddelbuettel2011}
Eddelbuettel, D. and Francois, R. (2011).
\newblock {Rcpp: Seamless R and C++ integration}.
\newblock {\em Journal of Statistical Software}, 40:1--18.

\bibitem[Fan et~al., 2006]{Fan2006}
Fan, Y., Brooks, S.~P., and Gelman, A. (2006).
\newblock {Output assessment for Monte Carlo simulations via the score
  statistic}.
\newblock {\em Journal of Computational and Graphical Statistics},
  15(1):178--206.

\bibitem[Faure and Lemieux, 2009]{Faure2009}
Faure, H. and Lemieux, C. (2009).
\newblock {Generalized Halton sequences in 2008: a comparative study}.
\newblock {\em ACM Transactions on Modeling and Computer Simulation},
  19:15:1--15:31.

\bibitem[Flegal et~al., 2008]{flegal2008markov}
Flegal, J.~M., Haran, M., and Jones, G.~L. (2008).
\newblock Markov chain monte carlo: Can we trust the third significant figure?
\newblock {\em Statistical Science}, 23(2):250--260.

\bibitem[Flegal and Jones, 2011]{Flegal2011}
Flegal, J.~M. and Jones, G.~L. (2011).
\newblock {Implementing MCMC: Estimating with confidence}.
\newblock In {\em {Handbook of Markov Chain Monte Carlo}}, pages 175--197.
  Chapman and Hall/CRC.

\bibitem[Fung et~al., 2019]{fung2019mpcmp}
Fung, H.-T., Alwan, A., Wishart, J., and Huang, A. (2019).
\newblock {mpcmp: Mean-parametrized Conway--Maxwell--Poisson regression}.

\bibitem[Gelman and Rubin, 1992]{gelman1992inference}
Gelman, A. and Rubin, D.~B. (1992).
\newblock Inference from iterative simulation using multiple sequences.
\newblock {\em Statistical science}, pages 457--472.

\bibitem[Geweke et~al., 1991]{geweke1991evaluating}
Geweke, J.~F. et~al. (1991).
\newblock Evaluating the accuracy of sampling-based approaches to the
  calculation of posterior moments.
\newblock Technical report, Federal Reserve Bank of Minneapolis.

\bibitem[Geyer, 1992]{geyer1992practical}
Geyer, C.~J. (1992).
\newblock Practical markov chain monte carlo.
\newblock {\em Statistical science}, pages 473--483.

\bibitem[Goldstein et~al., 2015]{Goldstein2015}
Goldstein, J., Haran, M., Simeonov, I., Fricks, J., and Chiaromonte, F. (2015).
\newblock {An attraction-repulsion point process model for respiratory
  syncytial virus infections}.
\newblock {\em Biometrics}, 71(2):376--385.

\bibitem[Gorham and Mackey, 2015]{Gorham2015}
Gorham, J. and Mackey, L. (2015).
\newblock {Measuring sample quality with Stein's method}.
\newblock In {\em Advances in Neural Information Processing Systems},
  volume~28.

\bibitem[Gorham and Mackey, 2017]{Gorham2017}
Gorham, J. and Mackey, L. (2017).
\newblock {Measuring sample quality with kernels}.
\newblock In {\em Proceedings of the 34th International Conference on Machine
  Learning}, pages 1292--1301.

\bibitem[Guikema and Goffelt, 2008]{Guikema2008}
Guikema, S.~D. and Goffelt, J.~P. (2008).
\newblock {A flexible count data regression model for risk analysis}.
\newblock {\em Risk Analysis}, 28:213--223.

\bibitem[Hastings, 1970]{Hastings1970}
Hastings, W.~K. (1970).
\newblock {Monte Carlo sampling methods using Markov chains and their
  applications}.
\newblock {\em Biometrika}, 57(1):97--109.

\bibitem[Huang, 2017]{Huang2017}
Huang, A. (2017).
\newblock {Mean-parametrized Conway–Maxwell–Poisson regression models for
  dispersed counts}.
\newblock {\em Statistical Modelling}, 17:359--380.

\bibitem[Hughes et~al., 2011]{Hughes2011autologistic}
Hughes, J., Haran, M., and Caragea, P.~C. (2011).
\newblock {Autologistic models for binary data on a lattice}.
\newblock {\em Environmetrics}, 22(7):857--871.

\bibitem[Hunter, 2007]{Hunter2007}
Hunter, D.~R. (2007).
\newblock {Curved exponential family models for social networks}.
\newblock {\em Social Networks}, 29(2):216--230.

\bibitem[Hunter and Handcock, 2006]{Hunter2006}
Hunter, D.~R. and Handcock, M.~S. (2006).
\newblock {Inference in curved exponential family models for networks}.
\newblock {\em Journal of Computational and Graphical Statistics},
  15(3):565--583.

\bibitem[Hunter et~al., 2008]{Hunter2008}
Hunter, D.~R., Handcock, M.~S., Butts, C.~T., Goodreau, S.~M., and Morris, M.
  (2008).
\newblock {ergm: A package to fit, simulate and diagnose exponential-family
  models for networks}.
\newblock {\em Journal of Statistical Software}, 24(3):1--29.

\bibitem[Ihaka and Gentleman, 1996]{Ihaka1996}
Ihaka, R. and Gentleman, R. (1996).
\newblock {R: A language for data analysis and graphics}.
\newblock {\em Journal of Computational and Graphical Statistics},
  5(3):299--314.

\bibitem[Ising, 1925]{ising1925}
Ising, E. (1925).
\newblock {Beitrag zur theorie des ferromagnetismus}.
\newblock {\em Zeitschrift f{\"{u}}rPhysik A Hadrons and Nuclei},
  31(1):253--258.

\bibitem[Jones et~al., 2006]{Jones2006}
Jones, G.~L., Haran, M., Caffo, B.~S., and Neath, R. (2006).
\newblock {Fixed-width output analysis for Markov chain Monte Carlo}.
\newblock {\em Journal of the American Statistical Association},
  101:1537--1547.

\bibitem[Kass et~al., 1998]{kass1998}
Kass, R.~E., Carlin, B.~P., Gelman, A., and Neal, R.~M. (1998).
\newblock {Markov Chain Monte Carlo in Practice: A Roundtable Discussion}.
\newblock {\em The American Statistician}, 52(2):93--100.

\bibitem[Lauritzen and Wermuth, 1989]{Lauritzen1989}
Lauritzen, S.~L. and Wermuth, N. (1989).
\newblock {Graphical models for associations between variables, some of which
  are qualitative and some quantitative}.
\newblock {\em The Annals of Statistics}, 17:31--57.

\bibitem[Lee and Hastie, 2015]{Lee2015}
Lee, J.~D. and Hastie, T.~J. (2015).
\newblock {Learning the structure of mixed graphical models}.
\newblock {\em Journal of Computational and Graphical Statistics}, 24:230--253.

\bibitem[Lee et~al., 2019]{Lee2019}
Lee, J.~E., Nicholls, G.~K., and Ryder, R.~J. (2019).
\newblock {Calibration Procedures for Approximate Bayesian Credible Sets}.
\newblock {\em Bayesian Analysis}, 14(4):1245--1269.

\bibitem[Lenz, 1920]{lenz1920}
Lenz, W. (1920).
\newblock {Beitrag zum Verst{\"{a}}ndnis der magnetischen Erscheinungen in
  festen K{\"{o}}rpern}.
\newblock {\em Physikalische Zeitschrift}, 21:613--615.

\bibitem[Leucht and Neumann, 2013]{Leucht2013}
Leucht, A. and Neumann, M.~H. (2013).
\newblock {Dependent wild bootstrap for degenerate U- and V-statistics}.
\newblock {\em Journal of Multivariate Analysis}, 117:257--280.

\bibitem[Liang, 2010]{Liang2010}
Liang, F. (2010).
\newblock {A double Metropolis-Hastings sampler for spatial models with
  intractable normalizing constants}.
\newblock {\em Journal of Statistical Computation and Simulation},
  80(9):1007--1022.

\bibitem[Liang et~al., 2016]{Liang2016}
Liang, F., Jin, I.~H., Song, Q., and Liu, J.~S. (2016).
\newblock {An adaptive exchange algorithm for sampling from distributions with
  intractable normalizing constants}.
\newblock {\em Journal of the American Statistical Association},
  111(513):377--393.

\bibitem[Lyne et~al., 2015]{Lyne2015}
Lyne, A.~M., Girolami, M., Atchad{\'{e}}, Y., Strathmann, H., and Simpson, D.
  (2015).
\newblock {On Russian Roulette estimates for Bayesian inference with
  doubly-intractable likelihoods}.
\newblock {\em Statistical Science}, 30(4):443--467.

\bibitem[Metropolis et~al., 1953]{Metropolis1953}
Metropolis, N., Rosenbluth, A.~W., Rosenbluth, M.~N., Teller, A.~H., and
  Teller, E. (1953).
\newblock {Equation of state calculations by fast computing machines}.
\newblock {\em The journal of chemical physics}, 21(6):1087--1092.

\bibitem[M{\o}ller et~al., 2006]{moller2006}
M{\o}ller, J., Pettitt, A.~N., Reeves, R., and Berthelsen, K.~K. (2006).
\newblock {An efficient Markov chain Monte Carlo method for distributions with
  intractable normalising constants}.
\newblock {\em Biometrika}, 93(2):451--458.

\bibitem[M{\"{u}}ller, 1997]{Muller1997}
M{\"{u}}ller, A. (1997).
\newblock {Integral probability metrics and their generating classes of
  functions}.
\newblock {\em Advances in Applied Probability}, 29(2):429--443.

\bibitem[Murray et~al., 2006]{murray2006}
Murray, I., Ghahramani, Z., and Mackay, D. J.~C. (2006).
\newblock {MCMC for doubly-intractable distributions}.
\newblock In {\em Proceedings of the 22nd Annual Conference on Uncertainty in
  Artificial Intelligence}, pages 359--366.

\bibitem[Park and Haran, 2018]{Park2018}
Park, J. and Haran, M. (2018).
\newblock {Bayesian inference in the presence of intractable normalizing
  functions}.
\newblock {\em Journal of the American Statistical Association},
  113(523):1372--1390.

\bibitem[Park and Haran, 2020]{Park2020}
Park, J. and Haran, M. (2020).
\newblock {A function emulation approach for doubly intractable distributions}.
\newblock {\em Journal of Computational and Graphical Statistics},
  29(1):66--77.

\bibitem[Propp and Wilson, 1996]{propp1996}
Propp, J.~G. and Wilson, D.~B. (1996).
\newblock {Exact sampling with coupled Markov chains and applications to
  statistical mechanics}.
\newblock {\em Random Structures {\&} Algorithms}, 9(1-2):223--252.

\bibitem[Rao et~al., 2016]{Rao2016}
Rao, V., Lin, L., and Dunson, D.~B. (2016).
\newblock {Data augmentation for models based on rejection sampling}.
\newblock {\em Biometrika}, 103:319--335.

\bibitem[Ribeiro et~al., 2020]{Ribeiro2020}
Ribeiro, E.~E., Zeviani, W.~M., Bonat, W.~H., Demetrio, C.~G., and Hinde, J.
  (2020).
\newblock {Reparametrization of COM–Poisson regression models with
  applications in the analysis of experimental data}.
\newblock {\em Statistical Modelling}, 20:443--466.

\bibitem[Robert and Casella, 2004]{Robert2004}
Robert, C.~P. and Casella, G. (2004).
\newblock {\em Monte Carlo statistical methods}.
\newblock Springer.

\bibitem[Robins et~al., 2007]{Robins2007}
Robins, G., Pattison, P., Kalish, Y., and Lusher, D. (2007).
\newblock {An introduction to exponential random graph (p*) models for social
  networks}.
\newblock {\em Social Networks}, 29(2):173--191.

\bibitem[Roy, 2020]{Roy2020}
Roy, V. (2020).
\newblock {Convergence diagnostics for Markov chain Monte Carlo}.
\newblock {\em Annual Review of Statistics and Its Application}, 7:387--412.

\bibitem[S{\'a}ez-Castillo and Conde-S{\'a}nchez, 2013]{Saez-Castillo2013}
S{\'a}ez-Castillo, A.~J. and Conde-S{\'a}nchez, A. (2013).
\newblock {A hyper-Poisson regression model for overdispersed and
  underdispersed count data}.
\newblock {\em Computational Statistics and Data Analysis}, 61:148--157.

\bibitem[Shmueli et~al., 2005]{Shmueli2005}
Shmueli, G., Minka, T.~P., Kadane, J.~B., Borle, S., and Boatwright, P. (2005).
\newblock {A useful distribution for fitting discrete data: Revival of the
  Conway–Maxwell–Poisson distribution}.
\newblock {\em Journal of the Royal Statistical Society: Series C},
  54:127--142.

\bibitem[Stein, 1972]{Stein1972}
Stein, C. (1972).
\newblock {A bound for the error in the normal approximation to the
  distribution of a sum of dependent random variables}.
\newblock In {\em Proceedings of the Sixth Berkeley Symposium on Mathematical
  Statistics and Probability}, volume~2, pages 583--602.

\bibitem[Strauss, 1975]{strauss1975}
Strauss, D.~J. (1975).
\newblock {A model for clustering}.
\newblock {\em Biometrika}, 62(2):467--475.

\bibitem[Tan and Friel, 2020]{Tan2020}
Tan, L.~S. and Friel, N. (2020).
\newblock {Bayesian variational inference for exponential random graph models}.
\newblock {\em Journal of Computational and Graphical Statistics}, 29:910--928.

\bibitem[Vats et~al., 2019]{Vats2019}
Vats, D., Flegal, J.~M., and Jones, G.~L. (2019).
\newblock {Multivariate output analysis for Markov chain Monte Carlo}.
\newblock {\em Biometrika}, 106:321--337.

\bibitem[Vats et~al., 2022]{Vats2022}
Vats, D., Gon{\c{c}}alves, F., {\L}atuszy{\'n}ski, K., and Roberts, G. (2022).
\newblock {Efficient Bernoulli factory Markov chain Monte Carlo for intractable
  posteriors}.
\newblock {\em Biometrika}, 109:369--385.

\bibitem[Xing et~al., 2019]{Xing2019}
Xing, H., Nicholls, G.~K., and Lee, J.~E. (2019).
\newblock {Calibrated approximate Bayesian inference}.
\newblock In {\em Proceedings of the 36th International Conference on Machine
  Learning}, pages 6912--6920.

\end{thebibliography}

\end{document}